\documentclass[twocolumn,showpacs,preprintnumbers,amsmath,amssymb,superscriptaddress,nofootinbib]{revtex4-1}
\usepackage{graphicx,dcolumn,bm}
\usepackage{enumerate}
\usepackage[usenames,dvipsnames]{color}
\usepackage{mathrsfs}
\usepackage{hyperref}
\usepackage{comment}
\usepackage{enumitem}

 \usepackage[utf8]{inputenc}
\usepackage{dutchcal}

\DeclareMathOperator{\grad}{grad}
\let\div\undefined
\DeclareMathOperator{\div}{div}
\DeclareMathOperator{\vol}{vol}
\DeclareMathOperator{\const}{const}
\DeclareMathOperator{\Tr}{Tr}

\usepackage{accents}
\newlength{\dhatheight}
\newcommand{\doublehat}[1]{%
    \settoheight{\dhatheight}{\ensuremath{\widehat{#1}}}%
    \addtolength{\dhatheight}{-0.35ex}%
    \widehat{\vphantom{\rule{1pt}{\dhatheight}}%
    \smash{\widehat{#1}}}}
    
    \usepackage{accents}
\newlength{\dhuatheight}
\newcommand{\triplehat}[1]{%
    \settoheight{\dhuatheight}{\ensuremath{\widehat{#1}}}%
    \addtolength{\dhuatheight}{-0.35ex}%
    \doublehat{\vphantom{\rule{1pt}{\dhuatheight}}%
    \smash{\doublehat{#1}}}}
    
\begin{document}
\newcommand{\nino}[1]{\textcolor{BrickRed} {\bf#1}}
\newcommand{\shelly}[1]{\textcolor{blue} {#1}}
\newcommand{\detlef}[1]{\textcolor{Green} {#1}}

\newcommand{\psicap}{\Psi}
\newcommand{\rr}{\mathrm{L}}
\newcommand{\rrr}{\rho}
\newcommand{\Om}{{\boldsymbol{\Omega}}}

\newcommand{\g}{\mathsf{g}}
\newcommand{\CS}{\mathcal{Q}}
\newcommand{\CSO}{\QS}

\newcommand{\bigcdot}{\boldsymbol{\cdot}}
\newcommand{\q}{{\boldsymbol{q}}}
\newcommand{\C}{C}

\newcommand{\ve}{\boldsymbol{e}}

\newcommand{\qs}{q}

\newcommand{\QA}{\pmb{\mathcal{Q}}}
\newcommand{\QS}{\mathcal{Q}}

\newcommand{\bomega}{\boldsymbol{\omega}}
\newcommand{\Qs}{Q}

\newcommand{\X}{\boldsymbol{\mathsf{X}}}
\newcommand{\Y}{\boldsymbol{\mathsf{Y}}}

\newcommand{\nablab}{\boldsymbol{\nabla}}
\newcommand{\Q}{\boldsymbol{Q}}
\newcommand{\spot}{\mathscr{V}}

\newcommand{\new}[1]{\textcolor{NavyBlue} { #1}}

\newcommand{\vect}[1]{\vec{\mathsf{#1}}}

\newcommand{\Hab}{\mathsf{H}}
\newcommand{\bHab}{\doublehat{\Hab}}
\newcommand{\bHabbb}{\triplehat{\Hab}}
\newcommand{\bHabb}{\mathcal{H}}
\newcommand{\bHabh}{\widehat{\mathsf{H}}}

\newcommand{\hHab}{\widehat{\mathsf{H}}}
\newcommand{\Op}{\mathsf{L}}
\newcommand{\euDelta}{\mathsf{\Delta}}

\newcommand{\lDeltag}{\widehat{\Delta}_B}

\newcommand{\Sc}{Schr\"odinger }
\newcommand{\psibh}{\widehat{\Psi}}
\newcommand{\psib}{\doublehat{\Psi}}
\newcommand{\psibb}{\triplehat{\Psi}}
\newcommand{\Psiq}{\psi}
\newcommand{\hPsiq}{\widehat{\psi}}
\newcommand{\psiq}{\psi}
\newcommand{\F}{$\mathscr{F}$ }
\newcommand{\E}{$\mathscr{E}$ }
\newcommand{\muF}{$\mu_\mathscr{F}$ }
\newcommand{\z}[1]{\textbf{#1}}

\newcommand{\ovq}{\vect{\mathcal{q}}}
\newcommand{\ovqc}{\widetilde{\boldsymbol{{q}}}}

\newcommand{\lift}[1] {\boldsymbol{#1}}
\newcommand{\liftw}[1] {\widehat{#1}}

\newcommand{\jac}{\mathfrak{J}}

\newcommand{\whpsi}{\widehat{\; \Psi\;}}
\newcommand{\whpsic}{\widehat{\; \psi\;}}

\newcommand{\gaupsi}{\widehat{\Psi}}

\newcommand{\gauconpsi}{\widehat{\psi}}

\newcommand{\zerogaupsi}{\widehat{\Psi}_{S}}
\newcommand{\firstgaupsi}{\widehat{\Psi}_{1}}

\newcommand{\secondgaupsi}{\widehat{\Psi}_{2}}
\newcommand{\thirdgaupsi}{\widehat{\Psi}_{3}}

\newcommand{\zerogauH}{\bHabh_{S}}

\newcommand{\firstgauH}{\bHabh_{1}}
\newcommand{\secondgauH}{\bHabh_{2}}
\newcommand{\thirdgauH}{\bHabh_{3}}

\newcommand{\Ppath}{\mathbb{P}}
\newcommand{\Epath}{\mathbb{E}}

\newcommand{\Pconf}{{P}}

\renewcommand{\psibh}{\firstgaupsi}
 \renewcommand{\bHabbb}{\thirdgauH}
 \renewcommand{\psibb}{\thirdgaupsi}
\renewcommand{\bHabb}{\zerogauH} 
 
 \renewcommand{\psib}{\secondgaupsi}
 \renewcommand{\bHab}{\secondgauH}

\title{Quantum Motion on Shape Space and the Gauge Dependent Emergence of\\ Dynamics and Probability in Absolute Space and Time}

\author{Detlef D\"urr}
\affiliation{Mathematisches Institut der Universit\"{a}t
     M\"{u}nchen, Theresienstra{\ss}e 39, 80333 M\"{u}nchen, Germany}
\author{Sheldon Goldstein}
\affiliation{Departments of Mathematics and Physics, Hill Center, Rutgers University,
     110 Frelinghuysen Road, Piscataway, NJ 08854-8019, USA}
\author{Nino Zangh\'{\i}}
\affiliation{Dipartimento di Fisica, Universit\`a di Genova, \\
Via Dodecaneso 33, 16146 Genova, Italy\\
\&\ Istituto Nazionale di Fisica Nucleare (Sezione di Genova) 
}

\date{\today}

\begin{abstract}
Relational formulations of classical mechanics and gravity  have been developed  by Julian Barbour and collaborators. Crucial to these formulations is the notion of shape space. We indicate here that the metric structure of shape space allows one to straightforwardly define a quantum motion, a Bohmian mechanics, on shape space. We show how this motion gives rise to the more or less  familiar theory in absolute space and time. We find that free motion on shape space, when lifted to configuration space, becomes an interacting theory. {Many different lifts are possible corresponding in fact to different choices of gauges.} Taking the laws of Bohmian mechanics on shape space as physically fundamental, we show how the theory can be statistically analyzed by using conditional wave functions, for subsystems of the universe, represented in terms of absolute space and time.

$$
\text{\large \it Dedicated to Joel Lebowitz, an invaluable friend and colleague}
$$
\end{abstract}

\maketitle
\tableofcontents

\section{Introduction}

Julian Barbour and Bruno Bertotti, in a very
inspiring and influential paper published  at the beginning of the eighties \cite{Barbour:1982gha} (for a recent overview, see \cite{barbour2012shape}, \cite{mercati2018shape} and references therein, see also \cite{anderson2011problem}),
transformed a long standing philosophical controversy
about  the nature of space and time into a well-defined physical problem. The philosophical  issue dates back to the  dispute between Isaac Newton, who {favored} {and argued for the need of } an absolute theory of space and time,  and Gottfried Wilhelm Leibniz, who insisted upon a relational approach,  also defended by Ernst Mach in the 19th  century.  The physical problem put forward by Barbour and Bertotti can be explained  by means of a very elementary and simplified  model of the universe.

Suppose we are given the configuration of a universe of $N$  particles.
And suppose we translate every particle of the configuration in the same 
direction by the same
amount. From a physical point of view it seems rather natural to
take the relational point of view that the two
configurations of the universe so obtained are physically equivalent or
identical. Similarly for any rotation. Going one step further, one regards two
configurations of the universe differing  only  by a dilation, i.e. by a uniform
expansion or contraction, as representing  in fact the same physical state of
the universe. 
The   space of all genuinely  physically different possible
 configurations so obtained---taking into account translations, rotations,  and
dilations---is  usually called {\it shape-space}.  
The name shape-space is  indeed natural: only the shape of a configuration of
particles is relevant, not its position or orientation  or overall size.

Given a kinematics   based on  shapes,   the next question to be addressed is that of their dynamics. 
In their seminal paper,  Barbour and Bertotti  proposed a dynamical principle based on what they called the intrinsic derivative and Barbour now calls {\em best matching}, which allows one to compare two shapes 
intrinsically, without any reference to the external space in which the particles  are embedded.  While the intrinsic comparison of shapes is compatible with positing  an  absolute Newtonian  time as in classical mechanics, it   naturally leads to  a relational notion of  time in which global changes of speed of the history of the universe give physically equivalent representations. Then the  dynamics can be reduced to geometry in the following sense:  a history of the universe is just a curve in shape space without any reference to a special parametrization of the curve  given by  absolute  Newtonian  time. 

The goal of the present paper is to extend the foregoing to the quantum case. We shall do this by 
considering the toy model  mentioned above in which the universe 
is modelled as an $N$-particle system. This will suffice to highlight the general feature of a relational quantum theory of the universe. However, we shall do so
 not by appealing to standard quantization schemes (see e.g., \cite{anderson2012problem}, \cite{PhysRevD.81.044035}), but by relying on the precise formulation of quantum theory provided by Bohmian mechanics \cite{bohm1952suggested,bell1987speakable,durr1992quantum,durr2012quantum,bricmont2016}.  Steps in this direction have been taken by Vassallo and Ip  \cite{vassallo2016conceptual}  and by Koslowski \cite{koslowski2017quantum}.

Bohmian mechanics  is a theory
providing a description of reality, compatible with all of the quantum
formalism, but free of any reference to observables or observers. In Bohmian mechanics a system of
particles is described in part by its wave function, evolving according to
Schr\"odinger's equation, the central equation of quantum theory. However,
the wave function provides only a partial description of the system. This
description is completed by the specification of the actual positions of
the particles. The latter evolve according to the ``guiding equation,''
which expresses the velocities of the particles in terms of the wave
function. Thus in Bohmian mechanics the configuration of a system of
particles evolves via a deterministic motion choreographed by the wave
function.  

Given the primary role of configurations, as opposed to operators and canonical quantization relations,  it should not come as a surprise 
that Bohmian mechanics can be very easily formulated on shape space: a wave function on shape space will govern the motion of a shape according to a guiding law analogous to the one of standard Bohmian mechanics. 
And to express the guiding law, as well as to write down Schr\"odinger's equation on shape space, 
all one needs is a {metric} on shape space.

Surprisingly (or maybe not),  the properties of metrics on shape space have been investigated  by applied mathematicians before the paper of Barbour and Bertotti, and for completly different reasons. What in physics is a configuration of $N$ particles, in statistics  is a set of data, and data analysis often requires that 
all information in a data set about its location, scale, and orientation
be  removed, so that the information that remains provides an intrinsic  description of the shape of the data. Indeed, the name ``shape space'' is due to the mathematicians that have been working on these problems of data analysis.  In particular David G. Kendall, whose early work on shape space dates back to the 1970s, was concerned with shape in archaeology and astronomy and also considered  the  motion of shapes formed by independent Brownian particles \cite{kendall1977diffusion}, while  
Fred Bookstein at about the same time began to study shape-theoretic problems in the particular context of zoology. 
Both recognized that the space of shapes can be represented by  Riemannian manifolds (see \cite{small2012statistical,kendall2009shape} for more background). We shall briefly review how to construct a metric on shape space  in Sect.~\ref{shapespace}.

Not only in Bohmian mechanics, but also in the classical theory of Barbour and Bertotti,  a metric on shape space plays a pivotal role in the formulation of the theory. Indeed, it turns out that Barbour's  best-matching principle is equivalent to a characterization of the dynamics as geodesic motion in shape space. Though this  fact was acknowledged by the authors in their original paper (and also in more recent publications by Barbour and collaborators), we think that sufficient emphasis has not been  given to it. 
Usually, classical  motion on shape space  is characterized by means  of Lagrangian or Hamiltonian formulations with constraints (see, e.g., \cite{barbour2003scale}).
While we agree that such methods of analytical mechanics  could be useful  in the analysis  of the  theory, we think that they obscure the geometrical structure  of the theory. So in Sect. \ref{motiononshapespace} we shall  provide a self-contained presentation of  the classical theory   by  emphasizing its geometrical content, in particular that the dynamics of shapes (even in presence of interactions) is geodesic motion on shape space. In Sect.  \ref{qmoss} we shall develop the Bohmian theory of motion and  highlight the similarities and differences between the classical case and the quantum case.

An important point  that we think has not been given sufficient emphasis is that the fundamental formulations of the theories---classical or quantum ---\emph{are} in shape space. And when the theories are formulated in shape space, one should consider first the simplest ones, namely the ``free'' theories  based only on the geometrical structures provided by the metric, without invoking any potential. This is in contrast with theories formulated in absolute space, for which free theories can't begin to account for the experimental data.  It is then natural to ask: when we represent the theories in absolute space, what form do the laws of motion  take? Is the representation unique or are there various representations yielding different looking laws of motion, some  unfamiliar and some more or less familiar? Moreover do  interacting theories emerge with nontrivial interactions, although in shape space the motion is free? 

To  answer  these questions it is helpful to represent absolute configuration space in geometrical terms as a fiber bundle, with shape space as base manifold and the fibers generated by the similarity group, i.e, by translations, rotations and dilations, which acting on configurations yields,  from a relational point of view,  physically equivalent states.  A representation in absolute configuration space of the motion in shape space is then given by  a ``lift'' of the motion from the base into  the fibers. 

Such lifts can rightly be called gauges. In the classical case it turns out that in some gauges the law looks unfamiliar but there is (at least) one gauge in which,  after performing a time change (representing indeed another gauge freedom when also time is seen as relational),  the law of motion is Newtonian with a potential appearing. The potential depends on the choice of the invariant metric (invariant under the action of the similarity group) in absolute configuration space, which we  introduce in Sect. \ref{shapespace}, where various possibilities for invariant metrics are given. 
The classical
 case is dealt with in Sect. \ref{emergencest}.

More or less the same is true for the quantum case,  where however the gauge yielding  ordinary Bohmian mechanics in absolute configuration space---which we call the Schr\"odinger gauge---emerges only for a stationary, i.e. time-independent, wave function (such as with the  Wheeler-DeWitt  equation) on shape space. This again is in line with regarding time as being relational, with an external absolute time playing no physical role.   

Also here, while the fundamental  physics is given by a free Bohmian dynamics in shape space, in the Schr\"odinger gauge potential terms appear. One potential term is determined by the scalar curvature induced by the  invariant metric on absolute configuration space. Another potential term arises from the gauge freedom we have to lift the Laplace-Beltrami operator from shape space to absolute configuration space, where an extra gauge freedom arises from allowing transformations  of the lifted wave function. To see the Schr\"odinger gauge arise, we invoke some mathematical facts from differential geometry. The details are in Sect. \ref{emergencestqu}.

Regarding  the motion in shape space as physically fundamental, we may well conclude from Sect.s \ref{emergencest} and \ref{emergencestqu} that the gauge freedom forces us to recognize that what we have  traditionally regarded as fundamental might in fact be imposed by us through our choice of gauge. This gauge freedom thus imparts a somewhat Kantian aspect to physical theory.

We next turn to the issue of probability, given by the quantum equilibrium  measure $|\Psi|^2$ on shape space. In assessing the relationship between probability on shape space and the usual Born-rule probabilities on absolute configuration space (associated with the natural lifts of the shape space dynamics to absolute space), we encounter several problems. First of all, since wave functions lifted from shape space are translation and scaling invariant, they fail to be normalizable. Another source of non-normalizability is  
  the  transition to the Schr\"odinger gauge. For  this gauge to be viable, as stated earlier, the wave function must be time-independent, and such wave functions typically fail, as with those of the Wheeler-DeWitt equation,  to be normalizable. 

So what could the associated non-normalizable  ``probabilities" physically mean? Moreover, the physical meaning of these measures would be obscure even if they were normalizable, since the absolute space degrees of freedom that transcend the relational ones are not observable, and the configuration $Q_t$ of the universe at ``time $t,$" whose distribution is supposed to be given by the Born rule, is, as we argue,  not physically meaningful. 

We address these questions in Sect. \ref{probonpaths}, in which we examine what {\it should} be physically and observationally meaningful, and find that the relevant probabilities for these are in fact given by a fundamental conditional probability formula (see \cite{durr1992quantum} for its meaning in the familiar Bohmian mechanics), as normalized conditional probabilities  arising from the non-normalizable quantum equilibrium measure on absolute configuration space. For this we use the notion of the wave function of a subsystem of shape space, a somewhat tricky business that is dealt with in Sect. \ref{ss}. 

We find in fact, somewhat to our surprise, that the non-normalizability of the wave function of the universe of quantum cosmology is, from a relational Bohmian perspective, a virtue rather than a vice.

\section{Shape Space}\label{shapespace}

\subsection{Shapes}
The totality of configurations $\q= (\vect{q}_1, \ldots , \vect{q}_N)$  of $N$ points  in Euclidean three-dimensional space 
forms the  configuration space $\QA= \{ \q\} =\mathbb{R}^{3N}$ of an $N$-particle system. We shall call 
$\QA$ the {\em absolute configuration space}.
On $\QA$ act  naturally the similarity transformations of Euclidean space, namely rotations, translations and dilations, since each of them acts naturally on each component of the configuration vector.
The totality of such transformations form the group $G$ of 
{\em similarity transformations} of Euclidean space.  Since the shape of a configuration is ``what is left'' when the effects associated with  rotations, translations and dilations are filtered away,
the totality of shapes, i.e.,   the {\em shape space}, is  the quotient space $\QS\equiv \QA/G$,  the set of equivalence classes
with respect to the  equivalence relations provided  by the similarity transformations of Euclidean space.

\begin{figure}[t!]
\centering
\includegraphics[width=.32\textwidth]{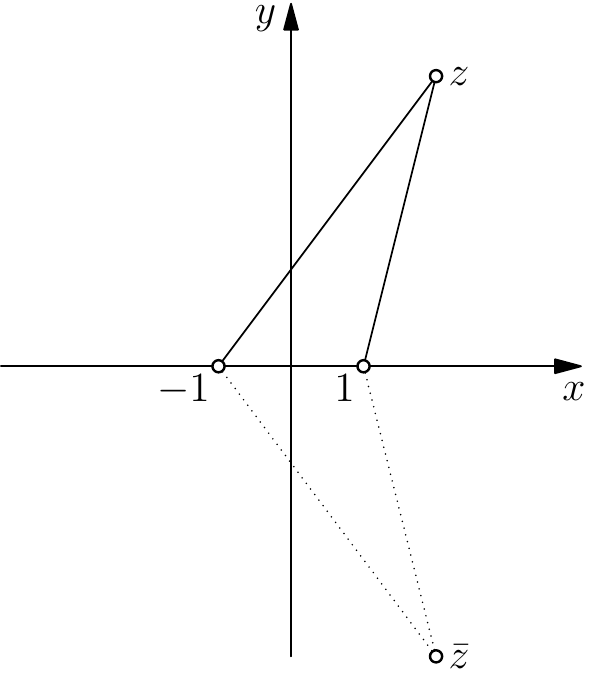}
\caption{\small  Representation of the shape space of  3 particles in terms of point $z$ in the complex upper half plane. Note that the complex conjugate $\bar{z}$ represents the same triangle since it can be obtained from  that of $z$ by a rotation in 3-dimensional Euclidean space.}
 \label{fig2}
\end{figure}

As such, shape space is not in general a manifold. To transform it into a manifold some massaging is needed (e.g., by excluding from $\QA$ coincidence points and collinear configurations), but we shall  not enter into this.\footnote{For more details on this issue, see, e.g., \cite{le1993riemannian} and reference therein.} Here, we shall assume that the appropriate massaging of  $\QA$ has been  performed  and that  $\QS$ is a manifold. Since the group of similarity transformations has dimension $7$ (3  for rotations + 3 for translations + 1 for dilations), the dimension of $\QS\equiv \QA/G$ is $3N- 7$. 

For $N=1$ and $N=2$  shape space is trivial (it contains just a single point).  $N=3$  corresponds to the  simplest not trivial shape space; it  has    dimension $3\times 3 - 7 =2$. 
It is worthwhile to give some details about this latter case. 
Three points in Euclidean space form a triangle,  so shape space is the space of all triangle shapes, with  ``triangle shape'' meaning now what is usually meant in elementary Euclidean geometry.  A nice representation of this space is in terms of points in the complex plane (called Bookstein-coordinates in \cite{small2012statistical}). On 
 the real axis, fix two points,  say $-1$ and $1$,  and put them in correspondence with  two vertices of the triangle. Then the third vertex is  in one-to-one correspondence with a complex number  in the upper half plane, as shown  in 
Fig. \ref{fig2}. Note  that the triangles in the lower half plane  are equivalent to those in the upper half plane by a suitable rotation in three dimensions. The real axis is the boundary of the manifold and its points represent degenerate collinear triangles. The point at infinity represents  the degenerate triangular shape with two coinciding  
vertices. So the space of triangle shapes (allowing two coincident vertices but not three) can be put in correspondence with the the {\em extended half upper complex plane}, which,  by stereographic projection, is topologically equivalent to a hemisphere. For $N>3$ the topological structure is more complicated (see, e.g.,   \cite{le1993riemannian}).

\subsection{Metrics on  Shape Space} 
Topology, of course, does not fix  a metric. A metric should provide  more, namely a natural notion of  distance on $\CS $. And since 
each point in $\CS $ represents a
class of configurations of $N$ particles related by a similarity transformation, the distance between two elements of $\CS $ induced by the
metric
should not recognize  any absolute configurational difference due to an overall
translation, or rotation, or dilation. In other words, it 
should  provide a measure of the \emph{intrinsic} difference
between two absolute configurations (that is, not involving any consideration
regarding
how such configurations are embedded in Euclidean  space).

Although the construction of such a metric   is well known in the mathematical literature   on random shapes \cite{le1993riemannian}, we prefer to give a self-contained  presentation  more suited for the physical applications. The bottom line is this: {\em a metric 
on  absolute configuration space  $ \QA$ that is invariant under the group $G$ of similarity transformations  of Euclidean space, given by  a suitable ``conformal factor'' (to be explained below), defines canonically a  metric on shape space $\QS$}. 

\begin{figure}[t!]
\centering
\includegraphics[width=.48\textwidth]{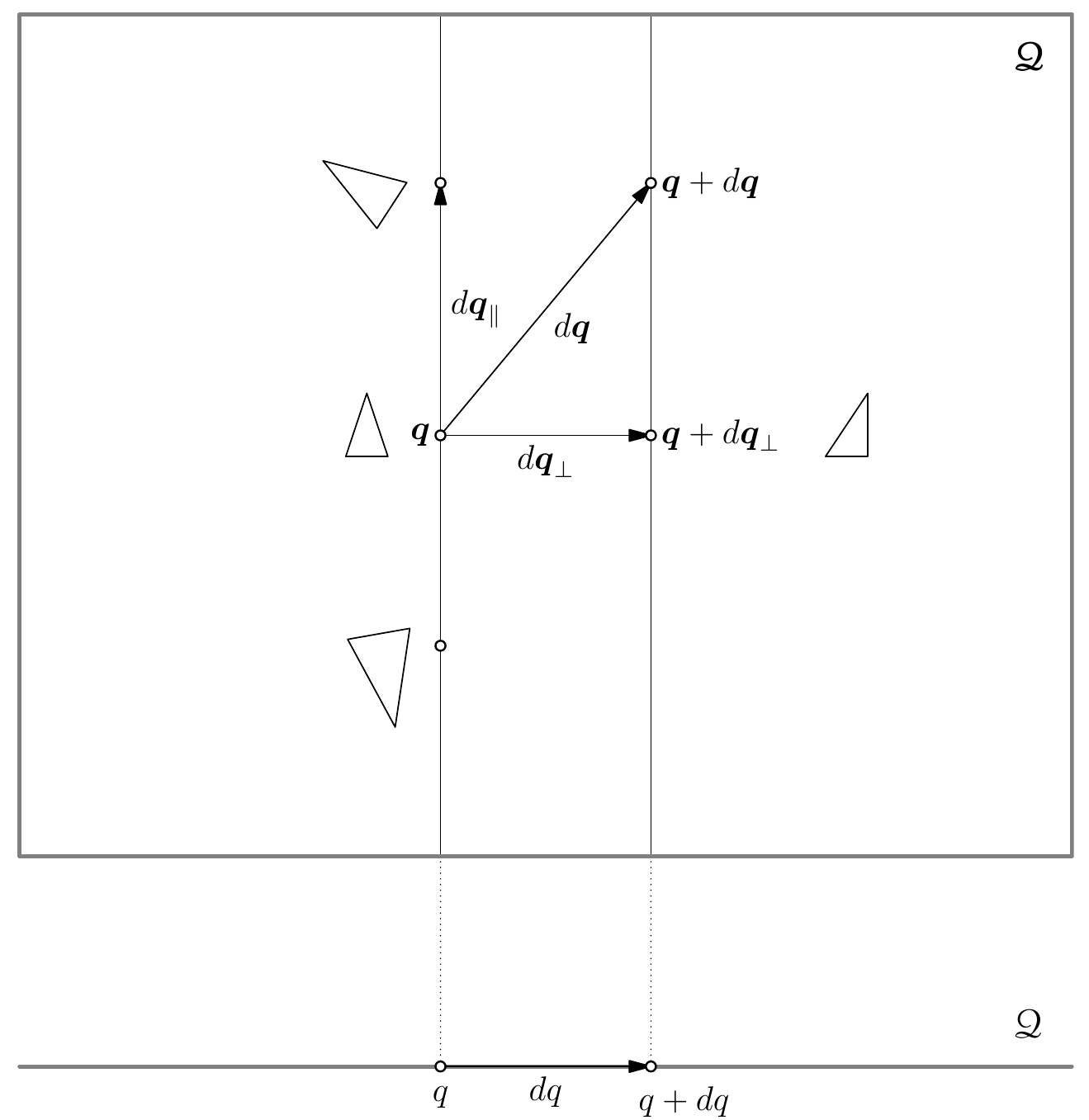}
\caption{\small Absolute configuration space $\QA$ and shape space $\QS$ (for a system of three particles). The fiber above shape $q$ consists of absolute configurations  differing by a similarity transformation of Euclidean  space and thus representing the same shape $q$. Real change of shape occurs only by a displacement to a  neighboring fiber 
 $q+dq$. Only the orthogonal component  $d{\q }_\perp$ of $d{\q }$ represents real change, while the vertical displacement $d{\q }_{\|}$ does not contribute;  ${\q }+ d{\q }_\perp$ is the absolute configuration in the fiber above $q + dq$   closest  to $q$ in the  sense of the $g_B$-distance (best matching). }
 \label{fig1}
\end{figure}

To understand why this is so,
observe first  that   absolute configuration space $ \QA$ can be regarded as  a fiber bundle  with  each fiber being  homeomorphic to $G$ and $\QS$ being its  base space (see  Fig. \ref{fig1}).   So, if $g$  is a metric invariant under any  element of $G$, the  tangent vectors at each point $\q\in\QA$    are   {\em naturally} split into ``vertical''   and ``horizontal,'' where by ``naturally'' we mean that the splitting itself is invariant under the action of $G$.  The vertical ones correspond to (infinitesimal) displacements along the fiber through $\q$ and the horizontal ones
are those that are orthogonal to the fiber, i.e., to the vertical ones, according to  the relation of orthogonality defined by $g$. More precisely, if $d\q$ is an infinitesimal displacement at $\q$, 
we have 
$$ d\q= d\q_{\|} + d\q_\perp\quad \text{with} \quad 
g (d\q_{\|}, d\q_\perp)  =0  $$ 
(see Fig. \ref{fig1}), with $ d\q_{\|} $ vertical and $d\q_\perp$ horizontal.

The corresponding Riemannian metric on $\QS$ is defined  as follows.  Let $\qs$ be a shape,  $\q$ be any absolute configuration in the fiber above $\qs$, and $d\q$ be any displacement at $\q$. Since $g$ is invariant under the group $G$,  the  length  of $d\q_\perp$  has the same value for all absolute configurations $\q$ above $\qs$.  Then we may set the   length  of $d\qs$   equal to that of $d\q_\perp$ and hence obtain the  Riemannian  metric $g_B$  on  $\CS$   
\begin{equation}
g_B  (d\qs, d\qs)  =   g  (d\q_\perp, d\q_\perp)    \,.
\label{eq:bmetr}
\end{equation}
The subscript $B$ stands for Barbour and Bertotti (as well as  base and best matching, see below).  

We shall now outline how to construct an invariant metric on $\QA$.   
 Let  $g_e$ be the mass-weighted Euclidean metric on $\QA$ with positive weights $m_ \alpha $, $ \alpha =1, \ldots, N$, 
(the masses of the particles),
in particle coordinates $\q = (\vect{q}_1, \ldots, \vect{q}_\alpha, \ldots, \vect{q}_N)$  given by 
 \begin{equation}
ds^2  =  \sum_{ \alpha =1}^N    m_\alpha d\vect{q}_ \alpha  \cdot   d \vect{q}_ \alpha   \,,
\label{eq:euclid}
\end{equation}
i.e., with 
$[g_e]_{ij} = m_{\alpha_i}  \delta_{ij}$,  where the $i$-th component 
 refers to the $\alpha_i$-th particle. The corresponding line element is
  \begin{equation}
|d\q |= \sqrt{ \sum_{ \alpha =1}^N    m_\alpha d\vect{q}_ \alpha  \cdot   d \vect{q}_ \alpha }  \,.
\label{eq:euclidmetric}
\end{equation}
%
The metric defined by \eqref{eq:euclid} is invariant under rotations and translations, but not under a dilation  $\q  \to\lambda \q $, where $\lambda$ is a positive constant. 
Invariance under dilations is achieved
by multiplying $|d\q|^2$ by a   scalar function $f(\q)$ that is invariant under rotations and translations and  is homogeneous   of degree $-2$. We call $f$ the {\em conformal factor}. So, for any choice of  $f$,   \begin{equation} 
g = f g_e\,, \quad\text{i.e,}\quad 
g (d\q, d\q) = f(\q) |d\q|^2 \,,
\label{eq:confmetric}
\end{equation}
is an invariant metric on $\QA$, yielding
the metric on  shape space
\begin{equation}
g_B(dq, dq)  =  f(\q)  |d\q_\perp |^2  \,.
\label{eq:bmetr1}
\end{equation}
For the associated  line element  we shall write
 \begin{align} \label{eq:bmetrxxx1}
 ds = |dq| =\sqrt{g_B(dq, dq)  } =\sqrt{f(\q)} \, |d\q_\perp |  \,. 
 \end{align}

\subsection{Best-Matching}
The distance on
$\CS$ induced
by $g_{B}$ is exactly the one  resulting from applying  Barbour's best
matching procedure. Consider two infinitesimally close shapes, $\qs$ and $\qs +
 d\qs$, and let $\q$ be any absolute representative of $\qs$, i.e., any point in
the fiber above $\qs$. The $g_B$-distance between these shapes is then given
by the $g$-length of the vector $d\q$ such that (i) $d\q$ is orthogonal to the
fiber above $\qs$ and  (ii) $\q + d\q$ is an absolute representative of $\qs +
d\qs$. It follows that $\q + d\q$ is the absolute configuration closest to $\q$ in
the fiber above $\qs +
 d\qs$. Thus the $g_B$-distance is the ``best matching'' distance.

\subsection{Conformal Factors}\label{conformalfactors}

Many choices of conformal factors  are possible. 
One that  was  originally suggested by
Barbour and Bertotti is\footnote{Here and in the following examples the conformal factors   are modulo   dimensional factors.}
\begin{align}f(\q) =f_a(\q)\equiv \left(\sum_{\alpha < \beta}
\frac{m_\alpha m_\beta}{|\vect{q}_{\alpha}-\vect{q}_{\beta}|}\right)^{2}\,. \end{align} 
Another example is \begin{align}  f(\q) =f_b(\q)\equiv {\rr ^{-2}} \,, \end{align} where
\begin{align}  \rr^2 &= \sum_\alpha m_\alpha{ \ovq_\alpha}^2 
=  \frac{1}{\sum_\alpha m_\alpha} \sum_{\alpha< \beta} m_\alpha m_\beta | 
\vect{q}_\alpha  - \vect{q}_\beta |^2  
 \label{mominsca}
\end{align}
with $ 
\ovq_\alpha = \vect{q}_\alpha - \vect{q}_\text{cm}\,,
$
the coordinates relative to the center of mass
\begin{align}
\vect{q}_\text{cm}  = \frac{\sum_\alpha m_\alpha\vect{q}_\alpha}{\sum_\alpha m_\alpha}\,.
 \end{align}
$I\equiv \rr^2 $ is sometimes called (but the terminology is not universal) the 
moment of inertia of the configuration $\q$  about  its center of mass. This quantity
 is half the trace of the moment of inertia tensor $\mathsf{M}$,
 \begin{align}   \rr^2= \frac12  \Tr  \mathsf{M} \,. \end{align}
 We recall that 
$ 
 \mathsf{M} = \mathsf{M} (\q) 
$, the  tensor  of inertia of the configuration $\q$ about any orthogonal cartesian system $x$,$y$,$z$  with origin  at the center of mass of the configuration $\q$,  has  matrix elements  given by the standard formula
\begin{align}\label{mateltensin}
M_{ij}= \sum_{\alpha=1}^N m_\alpha (\rrr_\alpha^2\delta_{ij} - \rrr_{\alpha i}  \rrr_{\alpha j}  )\,,
\end{align}
where $i,j=x,y,z$,  $\rrr_{\alpha x} \equiv  x_{\alpha} $,  $\rrr_{\alpha y} \equiv y_{\alpha} $, 
$\rrr_{\alpha z}  \equiv  z_{\alpha} $, and $\rrr_\alpha^2=  x_{\alpha}^2 +  y_{\alpha}^2 +z_\alpha^2$.

A  choice of conformal factor that has not been considered in the literature  is 
\begin{align}\label{pconfacty}
f(\q) =f_c(\q)\equiv   \rr^{-\frac{8}{7}} (\det\mathsf{M})^{-\frac{1}{7}}\,.
\end{align}
Since $\det\mathsf{M}$ scales as $\rr^6$, $f(\q)$ given  by \eqref{pconfacty}  scales as it should, namely, as  $\rr^{-2}$. 
Though at first glance this choice does not seem  natural, it is in fact so natural---once the motion of shapes  is analyzed from a quantum perspective, see  Sect. \ref{The Canonical Conformal Factor}---that  we shall  call  $f_c$
 the {\em canonical conformal factor}.

Finally, we  give other   two examples:
\begin{align}\label{cfd}
f(\q) &=f_d(\q) \equiv  \sum_{\alpha < \beta}
\frac{m_\alpha m_\beta}{|\vect{q}_{\alpha}-\vect{q}_{\beta}|^2} \\
\label{cfg}
f(\q) &=f_g(\q) \equiv \rr^{-1} \sum_{\alpha < \beta}  \frac{m_\alpha m_\beta}{|\vect{q}_{\alpha}-\vect{q}_{\beta}|} \,.
\end{align}
The first one corresponds to a natural modification of the Newtonian gravitational potential
and the second,  discussed in \cite{barbour2003scale}, corresponds to a dynamics very close to that of Newtonian gravity   (see Sect.  \ref{cmng} and Sect.  \ref{newgra}).


\section{Classical Motion on Shape Space}\label{motiononshapespace}

\subsection{Geodesic Motion}   The metric $g_{B}$ on shape space
directly yields a law of {\em free motion} on shape space, 
that is,
geodesic motion with constant speed. More explicitly, this is the motion $\Qs=\Qs(t)$ at constant speed  along the path  
that minimizes the length
 \begin{align}  \label{charterize1} 
  \int_{q_1}^{q_2} |dq| =   \int_{\q_1}^{\q_2}  \sqrt{ f(\q) }\, |d\q_\perp|
\end{align}
over all possible paths  connecting  two shapes $q_1$ and $q_2$ (if they are sufficiently close).
Note that the variational problem 
determines  only the path of the motion, but not the motion in time. 

Equivalently, a  geodesic motion    $\Qs=\Qs(t)$
is a motion  that parallel-transports its own tangent
vector, so 
\begin{align}\label{freemotsh}
D_{ \dot{Q}(t)}  \dot{Q}(t) = 0\,,
\end{align}
where $D_{ \dot{Q}(t)}$ is the covariant derivative with respect to the metric   $g_{B}$  along the curve $\Qs=\Qs(t)$.  Given the  initial conditions $Q(0)$ and $\dot{Q}(0)$, the motion will 
run at constant speed $v=| \dot{Q}(0)|$.

\subsection{Motion in a Potential}  Motion under the effect of the potential  $\spot= \spot(q)$  is  given by the obvious modification of \eqref{freemotsh}, namely Newton's equation
\begin{align}\label{Umotsh1}
D_{ \dot{\Qs}(t)}  \dot{\Qs}(t) = - \nabla_{g_{B}} \spot (\Qs)\,,
\end{align}
where $\nabla_{g_{B}} $ is the gradient with respect to the metric   $g_{B}$. This is equivalent to a characterization of the motion in terms of the Lagrangian
\begin{align}\label{Umotsh2}  L =\frac12   g_B\left(\frac{dq}{dt}, \frac{dq}{dt}   \right)-   \spot(q)= \frac12\left|\frac{dq}{dt}\right|^2- \spot(q).
\end{align}

\section{Quantum Motion on Shape Space}\label{qmoss}

\subsection{Bohmian Mechanics}

Various quantization schemes have been  put forward in order to provide a quantum theory of motion on shape space; for a thorough overview, see \cite{anderson2011problem}.  These schemes are mostly based on Dirac quantization of classical constrained systems or on Feynman path integration \cite{PhysRevD.81.044035}. We shall follow here a novel approach based on Bohmian mechanics.  Bohmian mechanics is  a completely
deter\-ministic---but distinctly non-Newtonian---theory of particles in
motion, with the wave function  itself guiding this motion. We shall explain below how this theory  can be naturally formulated on shape space, after a brief review of  the main features of the theory.

Bohmian mechanics is the minimal completion of Schr\"odinger's equation,
for a non-relativistic system of particles, to a theory describing a genuine
motion of particles. For Bohmian mechanics the state of a system of $N$
particles is described by its wave function $\psicap{} = \psicap{}(\vect{q}_1,
\dots ,\vect{q}_N) = \psicap{}(\q )$, a complex- (or spinor-) valued function on
the space of possible configurations $\q $ of the system, together with its
actual configuration $\Q$ defined by the actual positions $\vect{Q}_1,
\dots ,\vect{Q}_N$ of its particles. The theory is then defined by two
evolution laws. One is {\it Schr\"odinger's equation}
\begin{equation}
\label{Schroedinger}
  i\hbar \frac{\partial \psicap{}}{\partial t} = H\psicap{}\,,
\end{equation}
for $\psicap{}=\psicap{}_t$, the wave function at time $t$, where $H$  is the
non-relativistic (Schr\"odinger) Hamiltonian, containing the masses $m_k$, $k=1, \ldots, N$, of the
particles and a potential energy term $V$. For spinless particles, it is of the form
\begin{equation}
H=-\sum_{\alpha=1}^{N}
\frac{{\hbar}^{2}}{2m_{\alpha}}\vect{\nabla}^{2}_{\alpha} + V\,,
\label{sh}
\end{equation}
where $\vect{\nabla}_\alpha = \frac{\partial\,\;}{\partial \vect{q}_\alpha}$ is the gradient with respect to the position  of the $\alpha$-th particle.
The other law is the {\it the guiding law}, which,  for spinless particles, is given by the equation\footnote{\label{fn:proj}The general form of the guiding equation is 
\begin{equation*}\label{Bohm}
  \frac{d\vect{Q}_\alpha}{dt} = \frac{\hbar}{m_\alpha} \mathrm{Im} \frac{\psicap{}^*
  \vect{\nabla}_\alpha\psicap{}}{\psicap{}^* \psicap{}} ( \vect{Q}_1, \dots ,\vect{Q}_N )\,.
\end{equation*}  If $\psicap{}$ is
spinor-valued, the products in numerator and denominator should be
understood as scalar products. If external magnetic fields are present, the
gradient should be understood as the covariant derivative, involving the
vector potential.}
\begin{equation}
 \frac{d\vect{Q}_\alpha}{dt} = \frac{\hbar}{m_\alpha} \mathrm{ Im } \frac{  \vect{\nabla}_\alpha \psicap{}}{ \psicap{}} ( \vect{Q}_1, \dots ,\vect{Q}_N )
 \label{gl}
\end{equation}
for $\Q=\Q(t)$, the configuration at time $t$.  
For an $N$-particle system these two equations, together
with the detailed specification of the Hamiltonian $H$,  completely define the
Bohmian motion of the system. For sake of simplicity,  we shall consider  here just  Bohmian mechanics for spinless particles, with  Hamiltonian  \eqref{sh} and guiding law \eqref{gl}. For more details on the formulation of Bohmian mechanics for particles with spin or other internal degrees of freedom, see
\cite{durr2012quantum}.

While the {\it formulation\/}  of Bohmian mechanics does not involve the notion of
quantum observables, as given by self-adjoint operators---so that its
relationship to the quantum formalism  may at first appear somewhat obscure---it can in
fact be shown that  Bohmian mechanics not only accounts for quantum
phenomena, but also embodies
the
quantum formalism  itself as the very expression of its empirical import
\cite[Ch.2 and 3]{durr2012quantum}.

It is  worth noting that the guiding equation \eqref{gl} is intimately
connected with the de Broglie relation $\vect{p} = \hbar \vect{k}$,
proposed by de Broglie in late 1923, the consideration of which quickly led
Schr\"odinger to the discovery of his wave equation. The de Broglie relation connects a particle property, momentum
$\vect{p} = m\vect{v}$, to a wave property, the wave vector
$\vect{k}$ of a plane wave
$\psicap{}(\vect{q})=e^{i\vect{k}\cdot\vect{q}}$. From this one can easily
guess the guiding equation as the simplest possibility for an equation of
motion for $\Q$ for the case of a general wave function $\psicap{}$.

\subsection{Bohmian Motion on a Riemannian Manifold}

Note   that, given $V$,  the Bohmian mechanics defined by equations  \eqref{Schroedinger}, \eqref{sh}, and \eqref{gl} depends only upon the Riemannian structure $g=g_e$ given by \eqref{eq:euclid}.  In
terms of this Riemannian structure, the evolution equations \eqref{sh} and \eqref{gl}
 become
\begin{align}\label{eq:bmonmany1x}
\frac{dQ}{dt} &= \hbar  \, \mathrm{ Im } \frac{\nabla_g \psicap{}}{\psicap{}} \\
\label{eq:bmonmany2x}
i\hbar \frac{\partial \psicap{}}{\partial t} &= -\frac{\hbar^2}{2} \Delta_g \psicap{} + V\psicap{}\,,
\end{align}
where  $\Delta_g$ and  $\nabla_g$ are,
respectively, the Laplace-Beltrami operator and the gradient on the
configuration space equipped with this Riemannian structure. But there is nothing special about 
this particular Riemannian structure. Indeed, equations \eqref{eq:bmonmany1x} and 
 \eqref{eq:bmonmany2x}  as such hold  very generally  on {\em any} Riemannian manifold. 
 Thus, the formulation of a Bohmian dynamics on a Riemannian manifold requires only  as  basic 
ingredients  the differentiable and metric structure of the
manifold.

\subsection{Bohmian Motion on Shape Space}
Equations  \eqref{eq:bmonmany1x} and
 \eqref{eq:bmonmany2x} define  immediately  Bohmian motion on shape space with Riemannian metric $g= g_B$ as 
the motion on shape space 
given by  the evolution equations 
\begin{align}\label{eq:bmonmany1}
\frac{dQ}{dt} &= \hbar  \, \mathrm{ Im } \frac{\nabla_{B}  \psicap{}}{\psicap{}} \\
\label{eq:bmonmany2}
i\hbar \frac{\partial \psicap{}}{\partial t} &= -\frac{\hbar^2}{2} \Delta_B \psicap{} + \spot\psicap{}\,,
\end{align}
where  $\Delta_{B}$ and  $\nabla_{B} $ are,
respectively, the Laplace-Beltrami operator and the gradient on the
configuration space equipped with the Riemannian  metric \eqref{eq:bmetr1}.
This is all there is to say about the {\em formulation} of Bohmian mechanics on shape space. (This should   be contrasted with more involved  approaches as in, e.g.,  \cite{vassallo2016conceptual}.)

\section{The Emergence of Absolute Space and  Time in the Classical Case}\label{emergencest}

\subsection{Gauge Freedom in the Classical Case}
Given classical motion in shape space, there is a huge host of motions in absolute space that are compatible with it, the only constraint  being  that they should project down to free motion,  or the motion \eqref{Umotsh1}, in shape space.  This  freedom of choice is analogous to gauge freedom in gauge theories. Some choices are however more natural than others, as we shall discuss below. 

\subsection{Classical Motion in the Newton Gauge}\label{cmng}

A very natural  choice  of  a motion in  absolute configuration space is  the {\em horizontal lift}   of a  motion $\Qs=\Qs(t)$ in shape 
space, that is,  a motion $\Q= \Q(t)$ in absolute configuration space  that  starts at  
some  point $\q_1$ on the fiber above $\qs_1$  and is {\em horizontal}, i.e., the infinitesimal  displacements  $d\Q$ are all horizontal. (Note that  the final point $\q_2$ in the fiber above  $\qs_2$ is then uniquely determined.)  
 We call this choice  the {\em invariant gauge}.   
 
 We shall assume 
 $\spot=0$.\footnote{Our goal is to show that  the simplest dynamics  on shape space leads to a nontrivial dynamics  in a suitable gauge. The case  $\spot\neq 0$ will be considered in the next subsection.} Then  the   motion in the invariant gauge
 is geodesic motion with respect to the invariant metric (which explains the terminology).  To see this, observe that it follows from  \eqref{charterize1}   that the length of a horizontal lift of a path in shape space is  given by 
 \begin{align}\label{charterize2}  \int_{\q_1}^{\q_2}  \sqrt{ f(\q) }\,  |d\q_\perp|
= \int_{\q_1}^{\q_2}   \sqrt{ f(\q) }\, |d\q|      \,,  \end{align}
where the  equality follows from horizontality of the path. 
So, the path of  a horizontal lifted motion 
$\Q= \Q(t)$
has minimal length over all horizontal paths  connecting   $\q_1$ and $\q_2$, but   since any  non horizontal path  has a greater length,  $\Q(t)$  also minimizes the right hand side  of  \eqref{charterize2} 
over {\em all paths} connecting 
$\q_1$ and $\q_2$.  

We shall now show that by a suitable change of speed, we  get to another gauge that  we shall call the {\em Newton  gauge}, a gauge in which  the motion is  Newtonian, i.e., it satisfies Newton's equation $F=ma$ for suitable $F$.
To establish this, we first observe that the right hand side   of \eqref{charterize2} is of the form
\begin{align}\label{charterize77}
\int_{\q_1}^{\q_2} \sqrt{ E-V} |d\q |
\end{align}
for  $E=0$ 
and   $V(\q)= -  f (\q) $.
According to  the Jacobi principle,  
\eqref{charterize77} is minimized   
by 
 the path of  a Newtonian motion $\Q' = \Q'(t)$ in a potential  $V  $ and 
 total energy
\begin{align}\label{myspeed} E= \frac12 \left|\frac{d\Q'}{dt}\right| ^2 +V =0 \,. \end{align}
Thus the path of  a lifted motion  $\Q = \Q(t)$ is the same  as that of a Newtonian motion, but  its  speed along the path  is different:   
according to  \eqref {myspeed}    the speed  of the Newtonian motion is 
\begin{align}\label{charterize99} 
 \left|\frac{d\Q '}{dt}\right| = \sqrt{2(E-V)}= \sqrt{2 f}\,,
\end{align}
 while according to 
 \eqref{eq:bmetrxxx1}
the speed of   the lifted motion  is
\begin{align}\label{charterize199}   \left|\frac{d\Q }{dt}\right| = \frac{1}{\sqrt{f} }   \left|\frac{dq }{dt}\right| = \frac{v}{\sqrt{f}  }\, , \end{align}
with  $v$ the constant  speed of the motion on shape space.  So the two motions are different. But suppose
we allow for a change of  the flow of time and   replace $t$ with a new time variable $t'$ 
in such   a way that  the speed of the lifted motion  with respect to this new time  variable  equals the Newtonian speed $\sqrt{2f}$, 
\[   \left|\frac{d\Q }{dt'}    \right| =  \left|\frac{d\Q }{dt}   \frac{dt}{dt'} \right| = \sqrt{ 2f} \,, \]
whence, 
\begin{align}\label{timechange}
\frac{v}{\sqrt{f}  }  \frac{dt}{dt'} =  \sqrt{2f} \,, \quad\text{i.e.,} \quad   \frac{dt'}{dt} = \frac{v}{\sqrt{2} f} \,.
 \end{align}

Then $\Q= \Q(t')$, the lifted motion with respect to  this new time variable,  is indeed  a   Newtonian motion, that is, the particles positions $\vect{Q}_ \alpha$, $\alpha=1, \ldots, N$,  forming the configuration $\Q$ satisfy Newton's equations
\begin{align} \label{neteqab}m_\alpha \frac{d^2\vect{Q}_ \alpha }{d {t' }^2} = -\vect{\nabla}_\alpha V (\vect{Q}_ 1, \ldots \vect{Q}_ N )\,. \end{align}

One may wonder  about  the status of the time change \eqref{timechange}. If one considers time to be absolute,  $\Q= \Q(t)$ and $\Q'= \Q '(t )$ are two different motions. But if one takes 
a relational view about time, analogous to the relational view about space that we started with, 
$\Q= \Q(t)$ and $\Q'= \Q'(t )$ 
are the {\em same} motion. In other words, if time is relational,  changes of speed, such as that given by \eqref{timechange}, provide equivalent representations of the same motion. Accordingly, the use of one time variable instead of another is a matter of convenience, analogous to the choice of a gauge. 
The  choice of time variable for which Newton's equations \eqref{neteqab}  hold   is  
the gauge fixing condition that leads from the invariant gauge to  
 the Newton gauge; for the sake of simplicity, from now on we shall call it $t$ instead of $t'$.

The invariant gauge has been defined by requiring  that the path be horizontal.  It turns out that this is equivalent to the following conditions:
\begin{align}
&\sum_ {\alpha=1}^N  m_ \alpha  d\vect{Q}_ \alpha  = 0  \label{P=0}\\
&\sum_ {\alpha=1}^N  m_ \alpha \vect{Q}_ \alpha \times d\vect{Q}_ \alpha  = 0 \label{J=0} \\
&\sum_ {\alpha=1}^N  m_ \alpha \vect{Q}_ \alpha \cdot d\vect{Q}_ \alpha  = 0 \label{D=0} \,.
\end{align}
To see how this comes about, let  \begin{align}
\delta \vect{Q}_ \alpha  = \vect{\epsilon} + \vect{\theta}\times  \vect{Q}_ \alpha  + \lambda  \vect{Q}_ \alpha 
\end{align}
where $\vect{\epsilon}$,  $\vect{\theta}$, and $ \lambda$ are the infinitesimal parameters of a translation, a rotation and a dilation respectively, and let $  \delta \Q = (\delta\vect{Q}_1\, \ldots,\delta \vect{Q}_N) $. Then  \begin{align} \Q \to \Q+  \delta \Q \end{align}
is an infinitesimal  vertical transformation. Since the infinitesimal motion displacement $d\Q$ is purely horizontal, it must be orthogonal to $\delta \Q$, i.e.,
$
g(d\Q, \delta \Q) = 0 $,  which implies that 
\[ \vect{\epsilon}   \cdot \sum_ {\alpha=1}^N  m_ \alpha  d\vect{Q}_ \alpha  +
\vect{\theta} \cdot \sum_ {\alpha=1}^N  m_ \alpha  \vect{Q}_ \alpha  \times  d\vect{Q}_ \alpha 
+\lambda \sum_ {\alpha=1}^N  m_ \alpha  \vect{Q}_ \alpha  \cdot d\vect{Q}_ \alpha  = 0  \,.
\]
This  equality  is   satisfied  (for all $\epsilon$,  $\vect{\theta}$, and $\lambda$) only if  the terms  multiplying $\vect{\epsilon}$,  $\vect{\theta}$, and $ \lambda$ are separately zero,  whence \eqref{P=0}, \eqref{J=0}, and \eqref{D=0}.

The constraints  \eqref{P=0}, \eqref{J=0}, and \eqref{D=0} have a natural meaning for a theory aimed at describing the universe as a whole. So to speak, they minimize  the amount of motion when the universe is described in the invariant gauge. 

Moreover, the constraints  \eqref{P=0} and \eqref{D=0} are equivalent, respectively,  to the requirements that the motion $\mathbf{Q(}t)$ is such that the center of mass $(\sum m_\alpha)^{-1} \sum m_\alpha \vect{Q}_\alpha$ and the moment of inertia about the origin $\sum m_\alpha \vect{Q}_\alpha ^2 $ don't change. Clearly, these are natural gauge fixing choices corresponding to translational and dilational (scaling) symmetry. 
However, there can be no  function  on absolute configuration space which corresponds in a similar way to \eqref{J=0}. The constraint  \eqref{J=0} does not correspond to the constancy of a function on absolute configuration space.\footnote{This corresponds to the fact that the subspaces of the tangent spaces (at the points in  absolute configuration space) orthogonal to the fibers   don't correspond to a foliation of absolute configuration space into submanifolds orthogonal to the fibers. This is related to the fact that the curvature of the connection  relating the tangent spaces sitting at different points is non-vanishing and this, in its turn, is related to the Berry phase.\label{nontrivial}}

In the Newton gauge, \eqref{P=0}, \eqref{J=0}, and \eqref{D=0} can be expressed in terms of  
the familiar 
total momentum $\vect{P}$,  total angular momentum $ \vect {J}$ and (maybe less familiar)  dilational momentum $ D$ as
\begin{eqnarray}
 \vect{P}&=&\sum_{ \alpha =1}^N m_ \alpha \frac{d \vect{Q}_ \alpha }{d t} = 0\label{constraint1}\\
 \vect{J}&=&\sum_{ \alpha =1}^N  m_ \alpha  \vect{Q}_ \alpha \times \frac{d \vect{Q}_ \alpha }{d t} = 0\label{constraint2}\\
{D}&=&\sum_{ \alpha =1}^N  m_ \alpha  \vect{Q}_ \alpha \cdot \frac{d \vect{Q}_ \alpha }{d t} = 0\label{constraint3}\,.
\end{eqnarray}

\subsection{Some Remarks on Relational Space and Relational Time}\label{sec:potinshsp}
The first simple moral to draw form the foregoing is that 
free motion on shape space, i.e.,  for interaction energy  $\spot=0$, leads to  interaction energy $V\neq 0$  in the Newton gauge and so to an interacting particle dynamics 
in absolute spacetime (governed  by  Newton's laws \eqref{neteqab}).
 In other words, the geometry on shape space defined by the conformal factor $f$ manifests itself as potential energy $V$ among the particles in the Newton gauge. 

This remarkable fact is a direct consequence of  the  two main features of the theory under consideration.  One is our starting point, namely that shape space is fundamental, that is,  that {\em space is relational}. The other one    has emerged  in the analysis of   how shape dynamics appears in the Newton gauge:   motions following the same path  with different speeds
are indeed the {\em same} motion.   And this corresponds  to   {\em time being  relational}. 

This remarkable fact notwithstanding,
one may  still wonder what  sort of motion in absolute space corresponds to a shape dynamics with potential energy
$\spot\neq 0$.   
To answer to this question,  let us go back  to equations \eqref{Umotsh1} or \eqref{Umotsh2} defining 
interacting motion in shape space. Clearly, these equations are not in harmony with relational time: the acceleration in the LHS of \eqref{Umotsh1} or the Euler-Lagrange equations arising from \eqref{Umotsh2} rely on absolute time.  On the other hand, the characterization of motion in terms of 
the Jacobi  principle fits nicely with relational time. Adapted to the present case, this principle says 
that  the path followed by a motion in shape space is the path that minimizes
\begin{align}\label{chXXrYYze77}
\int_{\qs_1}^{\qs_2} \sqrt{ \mathscr{E} -\spot} |d\qs |\,,
\end{align}
where  $\mathscr{E}$ is any  given  fixed constant. And this  is in complete harmony with relational time: if time is relational all that matters is the path and not the speed along the path.  Note, however, that for  the relational dynamics defined by \eqref{chXXrYYze77}  changing the potential by adding a  constant changes the dynamics,  unlike the dynamics defined by  \eqref{Umotsh1} or \eqref{Umotsh2}.

Moreover,  if  interacting motion is defined according to \eqref{chXXrYYze77}, 
 it will still be   free  motion,  although  with respect to a different  metric:  the one defined by the conformal
 factor $ f_{\mathscr{E},\spot} = (\mathscr{E}  -\spot) f $ (with $\spot(\q) = \spot(q)$,  for any point $\q$ on the fiber above $q$).  As for the starting question concerning how the motion appears in the Newton gauge, the answer is rather obvious: just as above, but now for the conformal factor $  f_{\mathscr{E},\spot}= (\mathscr{E}  -\spot) f $.

The motion on shape space  characterized  by \eqref{chXXrYYze77} is defined  for any potential $\spot$ on  shape space; in particular,   it is defined for  
$\spot +  \mathscr{E}$.  So  the constant $\mathscr{E}$ can be absorbed in the potential;  that is, without  any loss  of generality, we may set $\mathscr{E}=0$ and consider only $  f_{\spot}=  -\spot f $. In this regard, it is important to observe 
that  changing the potential by a constant  changes the conformal factor and thus changes the dynamics. This is 
a peculiar aspect of  {\em relational mechanics}  (relational space and relational time), as opposed to the  usual Newtonian  mechanics (absolute space and absolute time), where a change of the potential by a constant does not change Newton's laws.

\subsection{Newtonian Gravitation}\label{newgra}
In the previous sections we found that in the Newton gauge, when the physical law on shape space is free motion (or even non-free motion), the potential $V=-f$ appears, where $f$ is the conformal factor. We mentioned some choices for $f$ in Sect.\ref{conformalfactors}. No such choices,  which are necessarily functions homogenous of degree $-2$, seem to
yield exactly the Newtonian gravitational potential $U_g$.  While we believe  the  detailed exploration of the implications of the models discussed here is worthwhile, we nonetheless  regard the models explored in this paper, both classical and quantum, as  toy models,  so that  such an analysis of them, with the expectation of recovering well established physics, might be somewhat inappropriate or premature.  

However,  it should be observed that some of the conformal factors given in Sect. \ref{conformalfactors}, e.g., $f_a$ and $f_g$,   indeed give rise to  a force law  in the Newton gauge that is very close to that of  the Newtonian gravitational potential.  Note for example  that  for  the conformal factor $f_g$ the corresponding potential   is of the form $V_g = \rr ^{-1} U_g$,   where $U_g$ is the  Newton gravitational potential and   $\rr $ in the Newton gauge is a constant of the motion. The 
force arising from this  potential adds to the  Newtonian force a very small  centripetal  correction that allows $I = \rr^2$,  the moment of  inertia  about  the center of mass,  to remain constant   \cite{barbour2003scale}.

\subsection{Gauge Freedom, Symmetry Breaking, and Newton's Bucket}\label{bucket}
The structures in an absolute space involved in the formulation of the geometry of shape space---in particular, the metric $g$ given by the conformal factor---are invariant under translations, rotations, and scaling. So, of course, is the classical dynamics on shape space,  since, by construction, translations, rotations, and scaling act trivially on shape space. The procedure defining the invariant gauge (Sect. \ref{cmng}) respects all of these symmetries. But scale invariance is broken in the Newton gauge because the time change \eqref{timechange}
 involved in the transition from the invariant gauge to the Newton gauge depends on the scale via $f$. This illustrates the obvious fact that the symmetries of the law of motion arising from the fundamental dynamics on shape space by a choice of gauge depends on the particular details defining that gauge.

A much larger class of symmetries for the shape space dynamics---also acting trivially---involves an independent group action $\g\in G$ at each ``time" (but not so independent that smoothness is lost). The most important and  familiar of these symmetries, when applied in a particular gauge, are uniformly growing translations (corresponding to Galilean boosts) and uniformly growing rotations (corresponding to the use of a rotating coordinate system or frame of reference). The former are a symmetry of the law of motion of the Newton gauge (ignoring the constraints \eqref{constraint1}-\eqref{constraint3}, which are obviously not preserved under boosts), since a change in position that depends linearly on time produces no change in the acceleration. The latter, however, is not a symmetry of the Newtonian law of motion.

The behavior of Newton's bucket, which has been used to argue against a relational understanding of space, is thus seen, in fact, to be a natural consequence of the relational view. That behavior is a consequence of Newtonian-like laws akin to those that emerge as the description in the Newton gauge of the fundamental dynamics on shape space. However, in the Newton gauge the total angular momentum of the universe must vanish, and this is incompatible with a (non-negligible) uniform rotation of the ``fixed stars." In a gauge corresponding to applying a uniformly growing rotation to the motion of the Newton gauge, the Newtonian law of motion is not obeyed, though the motion so obtained remains entirely compatible with the fundamental dynamics on shape space, a dynamics for which the behavior of the bucket depends essentially on its motion relative to that of the fixed stars.
\section{The Emergence of Absolute Space and  Time in the Quantum  Case}\label{emergencestqu}

\subsection{Gauge Freedom in the Quantum Case}
As in the classical case,  also the quantum theory is  about shapes, if one takes the standpoint of Bohmian mechanics. In this formulation of quantum mechanics, the role of the wave function is   that of  governing the motion of shapes. Moreover, as in the classical case,  there is {\em gauge freedom}: a huge host of motions in absolute space $\QA$  are compatible with Bohmian motion in shape space $\QS$. But now the presence of the wave function makes the freedom larger and subtler at the same time, as we shall explain in the following.

\subsection{The Schr\"odinger Gauge}\label{scg}
Let $\Qs=\Qs(t)$ be a Bohmian motion in shape space, that is,  a solution of \eqref{eq:bmonmany1} with the wave function  $\psicap{} $ being a solution of Schr\"odinger's  equation  \eqref{eq:bmonmany2} on shape space. For simplicity, we shall assume that  $\spot=0$ so that \eqref{eq:bmonmany2} becomes
\begin{align}\label{hamsf}  i\hbar \frac{\partial \Psiq}{\partial t} = \Hab \Psiq\,, \quad \Hab = -\frac{\hbar^2}{2} \Delta_B \end{align}
with $ \Delta_B $ the Laplace-Beltrami operator on shape space. 

As in the classical case, we wish  to characterize motions in absolute space that are compatible with motions
in shape space, that is,   motions  
$\Q =\Q(t)$ in $\QA$  that project down to $\Qs=\Qs(t)$ in $\QS$, i.e., such that
\begin{align}  \pi (\Q(t)) = \Qs(t) \,, \end{align}
where $\pi$ is the canonical projection from $\QA$ space to $\QS$.
Clearly, there  are  a great many possibilities  for compatible motions in absolute configuration space.

As in the classical case, one may restrict the possibilities by considering natural gauges. And as in the classical case where one  looks for gauges such that the absolute motions satisfy Newton's equations,
in the quantum case we now look for gauges such that the  compatible motions on  $\QA$ are themselves Bohmian motions, i.e. motions generated by  a  wave function in the usual sort of way.

For example, suppose that  we proceed as in the classical case and   take   a  horizontal lift    of a  motion $\Qs=\Qs(t)$ in shape 
space, that is,  an absolute  motion for which  the infinitesimal  displacements  $d\Q$ are all horizontal. Let us now consider the lift  to  $\QA$  of  a  wave function  $\psicap{}$  on $\QS$, namely, the wave function $ \firstgaupsi$ on absolute configuration space such that
\begin{equation}\label{strhl}  \firstgaupsi(\q) = \psicap{}(q)  \end{equation} for any point $\q$ on the fiber above $q$.
Let   $\nabla_g$be the gradient with respect to the invariant measure  \eqref{eq:confmetric}. Then
the vector
$\nabla_g  \psibh(\q)$ in $\QA$  is  horizontal and  the motions on $\QA$  defined by
\begin{align}\label{velogB}  \frac{d\Q}{dt}= \hbar  \, \mathrm{ Im } \frac{\nabla_{g} \psibh}{\psibh} \end{align}
are  horizontal lifts of motions on  $\QS$.
So, in the quantum case, horizontality is immediate.

Let us now consider the time evolution of the   lifted wave function $\psibh $ on  $\QA$.  Let  $\lDeltag$  be  a lift to absolute configuration space  of the Laplace-Beltrami operator $\Delta_B$ on shape space, namely an operator on $\QA$ such that
\begin{align}    \lDeltag \psibh  =  \Delta_B \psicap{}  \,. \end{align}
Then
\begin{align}\label{hamsfac}  i\hbar \frac{\partial\psibh}{\partial t} = \firstgauH \psibh\,,  \quad
\text{with}\quad
\firstgauH = -\frac{\hbar^2}{2}  \lDeltag  \,.\end{align}
It might seem   natural to guess that  $\lDeltag$   coincides  with $\Delta_{g}$,
the  Laplace-Beltrami operator  with respect to  $g$,   but this  is wrong;
nor is  $\bHabh_1$   a  familiar sort of Schr\"odinger Hamiltonian, with or without a potential term.

While  $\psibh $ need not obey any familiar Schr\"odinger-type equation, one may ask whether  there exists a gauge equivalent wave function that does. By gauge equivalent wave function we mean this:  If one writes  $\psibh $ as $Re^{(i/\hbar) S} $ one sees that the velocity field given by \eqref{velogB}  is just $\nabla_g S$, so transformations of the wave function
\begin{align}  \label{psiqtransf}  \psibh\to\psibh '  =  F\psibh\,, \end{align}
where $F$ is a positive function, do not change its phase and thus the velocity. 

It turns out that there exists a positive function  $F$ such that  $ \thirdgaupsi= F \firstgaupsi$ 
 (why  we use 3 here   instead of 2 
 will be clearer in Sect. \ref{sec:qmsgpr}) satisfies  a Schr\"odinger type  equation on absolute  configuration space for a suitable potential $V$, namely,  
\begin{align}\label{hamsf2}  i\hbar \frac{\partial \psibb}{\partial t} = \bHabbb \psibb   \end{align}
with
\begin{align} 
\label{eq:Steptwo} \bHabbb  
&= 
- \frac{\hbar^2}{2} \sum_{\alpha=1}^N  \vect{\nabla}_{\alpha}\cdot\frac{1}{ f m_\alpha }\vect{\nabla}_{\alpha} +V
\\
\label{eq:Steptwo}   &= -\frac{\hbar^2}{2} \nablab\bigcdot \frac{1}{f}\nablab +V\,,
\end{align}
%
%
where 
 $\nablab$ and 
 $\nablab\bigcdot$ are the  gradient and divergence with respect to the  mass-weighted Eucidean metric 
\eqref{eq:euclid}, i.e.,  
 \begin{align}\label{nsblab}  \nablab = \left(\frac{1}{m_1} \vect{\nabla}_{1} , \ldots,    \frac{1}{m_N} \vect{\nabla}_{N}    \right) \end{align}  
 and 
 \begin{align}\label{nsblab2}  \nablab \bigcdot= \left(\vect{\nabla}_{1} , \ldots,     \vect{\nabla}_{N}    \right) \bigcdot \, .\end{align} 
Here $f$ is the conformal factor, and
\begin{align}V &=V_1+V_2  \\
\label{V_1}
 V_1 &= 
-\frac{\hbar^2}{2}\frac{  \lDeltag J^{1/2}}{ J^{1/2}} \\
\label{exvtwo}
 V_2 &=   -\frac{\hbar^2}{2} f^{\frac{n}{4}}\Delta_g\big(f^{-\frac{n}{4}}\big)\,,
\end{align}
with
\begin{align}\label{forforJ}
J= \rr f^{7/2}  \sqrt{\det \mathsf{M}}\,,
\end{align}
where 
 $\rr=\rr(\q)$  is 
 given by  equation \eqref {mominsca} 
 and $\mathsf{M} = \mathsf{M} (\q) $  is the tensor  of inertia of the configuration $\q$ about any orthogonal cartesian system $x$,$y$,$z$  with origin in its  center of mass  and  with  matrix elements  given by   \eqref{mateltensin},  and where $ \lDeltag$ is the canonical lift \eqref{basicforJ}.

We shall now describe what we think is appropriate to be called the  {\em Schr\"odinger gauge}, the true quantum analogue of the Newton gauge.  If we take into account that time is relational, as we should, 
the fundamental equation for the wave function on shape space is  presumably the stationary equation
\begin{align}\label{eq:bmonmany3} 
-\frac{\hbar^2}{2} \Delta_B \psicap{} = \mathscr{E} \psicap{} \,,
\end{align}
where, for simplicity, we have set  $\spot=0$ and  $\mathscr{E}$ is any  given  fixed constant (for example $\mathscr{E}=0$). 

As before, let $\psibh$ be the lift of $\psicap{}$ to $\QA$, so that   $\psibh$ satisfies the equation
\begin{align}\label{hamsfacst}   (\bHabh_1 - \mathscr{E}) \psibh =0 \,, \end{align}
with $\bHabh_1$  a lift of $\Hab$ as in \eqref{hamsfac},
and  the evolution on the absolute configuration $\QA$  is  still given by \eqref{velogB}. 
But now, for relational time,  
{motions following the same path  with different speeds
are the same motion}. So, in the formula for the 
 gradient 
on the right hand side  of \eqref{velogB}, $\nabla_g = f^{-1}\nablab$, 
we may regard  $f$ as a change of speed  defining  a new time variable that for the  sake of simplicity we shall still call $t$ (``random time change''). Then
in absolute space  the guiding equation \eqref{velogB} becomes
\begin{align}\label{bohmabsolu}
\frac{d\vect{Q}_\alpha}{dt} =\frac{\hbar}{m_\alpha} \, \mathrm{ Im } \frac{\vect{\nabla}_{\alpha} \psibh}{\psibh} \,.
\end{align}



Again, $\psibh $ need not obey any familiar stationary Schr\"odinger-type equation. However, as before, we may exploit gauge freedom to transform \eqref{hamsfacst}  into a stationary Schr\"odinger-type equation. Indeed, we have an even greater gauge freedom in changing the wave function and the Hamiltonian, as will be shown below.
 In particular,  
there is a gauge, the Schr\"odinger gauge,  in which \eqref{eq:bmonmany3} becomes
\begin{align}\label{step31}
\bHabb \Phi= 0 
\end{align}
with 
\begin{align}\label{step312}
\bHabb= -\frac{\hbar^2}{2} \nablab^2 + U\,,
\end{align}
where $\nablab^2= \nablab\bigcdot \nablab$ is the  mass-weighed Euclidean Laplacian, and
\begin{align} \label{step32}
U= f (V_1  -\mathscr{E})   - \frac{\hbar^2}{8} \frac{n-2}{n-1} f  R_g  \,,
\end{align}
where $R_g$ is the scalar curvature of the invariant  metric~$g$.

\subsection{
Proofs of the Transitions to the Different Hamiltonians}\label{sec:qmsgpr} 
We shall now provide   proofs of the transitions from Hamiltonian $\bHabh_1$ in  equations \eqref{hamsfac} and \eqref{hamsfacst} to Hamiltonian  $\bHabh_3$, given by \eqref{eq:Steptwo},
in equation \eqref{hamsf2},  and Hamiltonian   $\bHabh_S$, given by \eqref{step312},  in  equation
 \eqref{step31}. 
The material presented here and in the following subsections is of a more mathematical character  and could be skipped in  first reading.

The lift  to $\QA$  of the Laplace-Beltrami operator $\Delta_B$ on  $\QS$ is by no means unique. There is however a ``canonical lift'' 
%
%
given by the formula
\begin{align}\label{basicforJ}
\lDeltag =  \jac  \,\div_g  \jac^{-1} \grad_g\,,
\end{align}
where  $\jac= \jac(\q)$  is a positive function on $\QA$ (unique up to a constant multiple), 
$\grad_g$  is  the gradient, given by (using the Einstein summation convention)
\begin{align}\label{gradcoo} \left( \grad_g\right)^i = g^{ij} \partial _j
\end{align}
in a coordinate basis $(\partial_1, \ldots, \partial_n)$,  and $ \div_g $ 
 is  the divergence whose action on a  vector field $\Y=(Y^1, \ldots, Y^n)$
in the coordinate basis $(\partial_1, \ldots, \partial_n)$ is 
\begin{align}\label{divcoo}  \div_g \Y = \frac{1}{\sqrt{|g|}}\partial_i \sqrt{|g|} Y^i  \,,\end{align}
where $|g| = |\det(g_{ij})| $ is the absolute value of the determinant of the metric tensor $g_{ij}$ in the given local coordinates.
The existence of a positive $\jac$ such that \eqref{basicforJ} defines a lift of $\Delta_B $  will be proven in  Sect.
\ref{derishjac} and formula \eqref{forforJ} for $\jac$  will be derived  in  Sect.\ref{compshjac}. We   call $\jac$ the {\em shape Jacobian}.

While $\lDeltag $ does not coincide with $\Delta_g$, the Laplace-Beltrami operator on  $\QA$, it is a minimal modification thereof. Just  compare  
\eqref{basicforJ} with  $\Delta_g$  as  ``div-grad'' operator, i.e.,
in local coordinates, 
\begin{align}
\label{eq:LB} 
\Delta_g = \div_g \grad_g &= \frac{1}{\sqrt{|g|}}\partial_i \sqrt{|g|}g^{ij}\partial_j \\ &=
 \frac{1}{f^{n/2}} \nablab \bigcdot f^{(n/2)-1} \nablab   \,,
 \label{eq:LB2} 
\end{align}
where in the second equality we have made explicit the invariant metric 
$g= f g_e$, with $g_e$ given by \eqref{eq:euclid},
in Euclidean particle coordinates: 
$g_{ij} =  f m_{\alpha_i}   \delta_{ij}$,
so that
$\sqrt{|g|}=  \left( m_1 \cdots  m_N \right)^{3/2}  f^{n/2}$, with $n=3N$,  and $g^{ij} =f^{-1} m_{\alpha_i}^{-1}  \delta_{ij}$.
Similarly,  
\begin{align}
\label{eq:LBJJ} 
\lDeltag  &=\frac{\jac}{\sqrt{|g|}}\partial_i \frac{\sqrt{|g|}}{\jac} g^{ij}\partial_j \\ &=  \frac{\jac}{ f^{n/2}} \nablab \bigcdot  \frac{f^{(n/2)-1}}{\jac} \nablab 
\,.
\end{align}

Note that $\Delta_g$ is self-adjoint on $L^2(d\mu_g)$,  the set of  functions 
on $\QA$  square integrable  with respect to the volume  element   defined  by the metric $g$,
\begin{equation} \label{dmugdef}
d\mu_g = \sqrt{|g|} dx_1 \cdots dx_n \propto  f^{n/2} d^3\vect{q}_1 \cdots d^3 \vect{q}_N\ \end{equation}
($n=3N$). In contrast, 
$ \lDeltag  $  is self adjoint  with   respect to the volume  element  $d\mu  =\jac^{-1} d\mu_g$. 

Let us  now consider the  effect of the gauge transformation \eqref{psiqtransf} on $\firstgauH = -   ({\hbar^2}/{2} )\lDeltag$.  Since  $ \psibh\to \psib= F\psibh $ is a  unitary  transformation \begin{align} \label{unitcond} U : L^2 (d\mu) \to L^2(F^{-2}d\mu)\,, \end{align} the  effect of   \eqref{psiqtransf}  is to  transform $\firstgauH $ into the unitarily
equivalent operator
\begin{align}  \label{htilde}   \bHab  =   U  \firstgauH  U^{-1} =  F \firstgauH  F^{-1} \,,\end{align}
so  that $\{ 
\firstgauH , \psibh\}$  and  $\{ 
 \bHab, \psib\}$  provide equivalent description of the dynamics.

A natural question  is   whether there is
an equivalent description   such that $ \bHab $
is  Schr\"odinger-like with  some potential. The key to   answering this question is  the following  theorem concerning second order partial  differential operators (see the Appendix for a proof): {\em Suppose $H_1$ and $H_2$ are second order partial  differential operators, both self-adjoint with respect to 
the same measure.  If they have the same pure 2nd derivative parts then
\begin{align} H_2 = H_1 + V\,. \end{align}
Moreover, if  $H_1  1 = 0$ (no constant part) then $V = H_2 1$.}

We  first apply this theorem to the operators  $H_1= -({\hbar^2}/{2} )\Delta_g$  and  $H_2= { \bHab}$, unitarily equivalent to ${ \firstgauH }$ according to \eqref{htilde}. 
Choosing $F= \jac^{-1/2}$ in  \eqref{htilde},  $H_2$  is self-adjoint with  respect to $\mu_g$. 
(According to \eqref{unitcond}, this operator is   self-adjoint with respect to $\jac d\mu = \jac\jac^{-1} d\mu_g= d\mu_g$.)
 So, $H_1$ and $H_2$  so defined  are  self-adjoint with respect to the same measure, have the same pure 2nd derivative parts (namely,
$- (\hbar^2/2) f^{-1} \nablab\bigcdot \nablab$)
  and $H_1  1 = 0$.  Thus, according to the theorem stated above,
\begin{align}\label{tildaH}
{ \bHab}= -\frac{\hbar^2}{2} \Delta_g + V_1  
\end{align}
with
\begin{align}\label{Vone}
 V_1= 
-\frac{\hbar^2}{2}\frac{  \lDeltag \jac^{1/2}}{ \jac^{1/2}} \,.
\end{align}
Let us now  perform a further transformation  on  ${\bHab}$   to make it   unitarily equivalent to the operator ${ \bHabbb} $ (see \eqref{eq:Steptwo}), which is self-adjoint with respect to the Lebesgue measure   $d^3\vect{q}_1 \cdots d^3 \vect{q}_N$.  Observing the form 
\eqref{eq:LB2}  of $\Delta_g $,   the  desired transformation is 
\begin{align}  \label{htilde2}   \bHab \to  \bHabbb    &=    f^{n/4}   \bHab  f^{-n/4} \\
&=  -\frac{\hbar^2}{2} f^{n/4}  \Delta_g  f^{-n/4} + V_1 \label{htilde3}\\
&\equiv H_2 + V_1 \label{htilde4}\,.
\end{align}

Consider now the operator  
\begin{align}\label{H-1}
H_1=  -\frac{\hbar^2}{2} \nablab \cdot \frac1f \nablab
\end{align} and note that $H_1$ and $H_2$
have the same pure 2nd derivative parts, are self-adjoint with respect to the same measure  (the Lebesgue measure) and $H_1  1 = 0$. Thus,
\begin{align}
H_2 =  H_1  +V_2
\end{align}
with 
 \begin{align}\label{exvtwos}
 V_2= H_21=  -\frac{\hbar^2}{2} f^{\frac{n}{4}}\Delta_g\big(f^{-\frac{n}{4}}\big)\,.
\end{align}
Finally, by  inserting $H_2$   into \eqref{htilde4}, we get
 \begin{align}\label{exvtwoxx}
  \bHabbb  =  H_1 + V_2 + V_1 = -\frac{\hbar^2}{2} \nablab\cdot \frac{1}{f}\nablab +V_1+ V_2\,,
 \end{align}
which is formula  \eqref{eq:Steptwo}  with
$V_1$ and  $V_2$ given by \eqref{V_1} and \eqref{exvtwo}.
 
Consider now the stationary equation corresponding to equation
\eqref{tildaH}, namely
\begin{align}\label{tildaHstat}  \left( { \bHab} - \mathscr{E}\right)    {\psib} = 0
\end{align}
and observe that now we may allow
a broader class of  transformations 
$\{ 
 { \bHab} ,  {\psib} \}   \to \{ 
 {\bHabb} ,  {\Phi} \}$ leading to an equivalent description of the dynamics. More precisely,  with a change  of $\psib$ according to \eqref{psiqtransf},
 \begin{align}  {\psib}    \to \zerogaupsi &= F  \psib \end{align}
 with $F >0$,  we need not demand now
 that the Hamiltonian gets transformed according to \eqref{htilde};
 the more general change
\begin{align} 
({ \bHab}   -\mathscr{E})   \to   \bHabb &= G ( { \bHab}  -\mathscr{E})  F^{-1}   \,, \end{align}
with
 $G > 0$   not necessarily equal to $F^{-1}$ suffices. 
 Recalling \eqref{tildaH}, we have
 \begin{align}
  \bHabb = H_2      + G F^{-1} ( V_1  - \mathscr{E})
  \,,\end{align}
with now $H_2$ defined as
  \[  H_2  =   -\frac{\hbar^2}{2} G   \Delta_g    F^{-1} \,.\]
Observing the form 
\eqref{eq:LB2}  of $\Delta_g $,  for the choice 
 \begin{align} F=  f^{\frac{n-2}{4} } \,, \quad  G =  f^{\frac{n}{2}}   f^{-\frac{n-2}{4} } = f^{\frac{n+2}{4}}  \,, \end{align}
$H_2$ has the same  pure 2nd derivative part as $H_1\equiv
 - ({\hbar^2}/{2} )   \Delta$;  moreover,   $H_1$ and $H_2$ so defined 
are self-adjoint with respect to   Lebesgue measure and $H_1  1 = 0$. Thus,
\[   H_2 =    -\frac{\hbar^2}{2} \nablab^2 + V_3\]
with  
\begin{align}V_3 =      -\frac{\hbar^2}{2}   f^{\frac{n+2}{4}}     \Delta_g  f^{-\frac{n-2}{4} }     \,.\end{align}

The potential $V_3 $  has a natural geometrical meaning. To see this, note that the scalar curvatures   $R_{g}$  and $R_{\tilde{g}}$ of 
two conformally related metrics  $g$ and
$\tilde{g} = \Lambda g$ are  related  by the formula  (see, e.g., \cite{marques2010scalar})
\[
R_{\tilde{g}}=\Lambda^{-1}\left(R_g -\frac{4(n-1)}{n-2} \Lambda^{-\frac{n-2}{4} }  \Delta _g  \Lambda^{\frac{n-2}{4}} \right)\,.
\]
Letting  $g$ be   the invariant metric  on $\QA$, 
$\tilde{g} $ be  the Euclidean metric on $\QA$ (so that $R_{\tilde{g}}= 0$) and $\Lambda = f^{-1}$, we obtain that
 \begin{equation}\label{curpot}
 R_g = \frac{4(n-1)}{n-2} f^{\frac{n-2}{4} }   \Delta _g  f^{-\frac{n-2}{4}} \,,
 \end{equation}
 whence 
\begin{align}
 V_3 =  -\frac{\hbar^2}{2} \frac{n-2}{4(n-1)} f  R_g\,.
\end{align}
Since $G F^{-1} = f$, we conclude that
\begin{align}
  \bHabb =  -\frac{\hbar^2}{2} \nablab^2 + f (V_1  -\mathscr{E})   - \frac{\hbar^2}{8} \frac{n-2}{n-1} f  R_g \,,\end{align}
 which coincides with  \eqref{step312}  for $U$ given  by \eqref{step32}.
 This completes the proofs of the transitions to the different Hamiltonians. 
 
Note that while  $V_2$  and $V_3$  do not depend on  the shape Jacobian $\jac$,  $V_1$  does. So, to find  $V_1$, we have first to find an explicit formula for $\jac$. This will be done in Sect. \eqref{derishjac}  and Sect. \eqref{compshjac}.

\subsection{Remarks on the  Bohmian Motion  
 in the Various Gauges}\label{rmkbm}
 
We have already   stated that the usual Bohmian motion  associated with 
$ \bHab$  is given by  \eqref{velogB}, see \eqref{eq:bmonmany1x}. This  is true also for $ \firstgauH $ and $\bHabbb$.  To see this, consider a Hamiltonian of the form \begin{equation}\label{genhamfg} H= -\frac{h^2}{2} F \nabla  \cdot  G \nabla   + U  \,, \end{equation}with $U$ a potential (multiplication operator), where $\nabla $ and $\nabla \cdot$
are, respectively,  the gradient and the divergence with respect to a metric with associated volume element $d\mu$. Then $H$ 
is  self-adjoint with respect to $F^{-1} d\mu$
and the velocity field generated by a solution $\widehat{\Psi}$ of  the Schr\"odinger equation associated with $H$  is 
\begin{equation} \label{eq:vgenea}
v =  \hbar  \, \mathrm{ Im } \frac{FG \nabla  \widehat{\Psi}}{\widehat{\Psi}} \,. 
\end{equation}  
Thus, in the ``1-gauge'' for $ \firstgauH $  involving $\lDeltag$ given by \eqref{basicforJ}, the Bohmian velocity \eqref{eq:vgenea} is
indeed \eqref{velogB} (with $\widehat{\Psi} =\firstgaupsi$), since in this gauge $\nabla= \grad_g$,  $\nabla\cdot = \div_g$, $F= \jac$ and $ G= \jac^{-1}$.  The   same velocity arises  in the  ``2-gauge''    \eqref{tildaH}  (with $\widehat{\Psi}=\widehat{\Psi}_2$),  since now  $F= G =1$ (and  $\nabla= \grad_g$,   $\nabla\cdot = \div_g$, as before), as well as 
 in the   ``3-gauge''   \eqref{exvtwoxx}   (with $\widehat{\Psi}=\thirdgaupsi$), since now $\nabla  = \nablab$, $\nabla \cdot = \nablab\bigcdot$, the usual    divergence,  and $F=1$, $G= 1/f$, so that \eqref{eq:vgenea}  equals \eqref{velogB} since $\nabla_g = f^{-1} \nablab$. On the other hand,   in the ``$S$-gauge''   for 
 $  \bHabb $ (the 
 Schr\"odinger gauge) the  Bohmian velocity is given by the usual formula  \eqref{bohmabsolu},
 which, as already stated, arises from \eqref{velogB} after a time change.
 

We shall address the status of probability measures in Bohmian mechanics on shape space and in the various gauges in Sect. \ref{probonpaths}. Here we shall just state some mathematical facts about the   quantum equilibrium measures (Born's rule) associated with the Bohmian motions in the various gauges. These measures are 
the same in all the first three gauges, though they assume different forms. In each gauge, they are the quantum equilibrium measure associated with the solutions
of the  wave equation in that gauge.

By construction, and more explicitly, 
since \eqref{genhamfg} is self-adjoint with respect to  $F^{-1} d\mu$,
$ \firstgauH $, $ \bHab$, and $\bHabbb $  are   self-adjoint with respect to 
 the measures $\jac^{-1} d\mu_g$, $d\mu_g$, and $d\lift{q}$ (the Lebesgue measure on $\QA$), respectively. Thus, the corresponding quantum equilibrium measures are, respectively,

  \begin{align}  \label{born1}
\left.
 \begin{aligned} 
&   |\firstgaupsi|^2  \jac^{-1} d\mu_g\\
 &   |\secondgaupsi|^2   d\mu_g\\
&   |\thirdgaupsi|^2 d\lift{q} 
 \end{aligned} \right\}  \equiv 
 d\mu^{\widehat{\Psi}}\,.
 \end{align}
The equality of these measures (up to a constant multiple) readily follows from  the relations between  the various gauges,
\begin{align}
\psibh(\q) &= \psicap{}(q) \,,\\
\secondgaupsi&= \jac^{-1/2} \psibh\,,\\
\thirdgaupsi &= f^{n/4}  \secondgaupsi\,,
\end{align}
and formula \eqref{dmugdef} for $d\mu_g$.  We finally note that in the
 Schr\"odinger gauge, as for the velocity, Born's probability law turns out to be the familiar one, namely,  
 \begin{equation}\label{muphi}
 d\mu^{\zerogaupsi} = |\zerogaupsi|^2   d\lift{q}\,.
 \end{equation}
 Note that 
 in going from the 3-gauge to the  $S$-gauge
 there  is no change of measure for self-adjointness of the Hamiltonian, so that   the change  $\thirdgaupsi\to \zerogaupsi$ leads in this case to a change in measure,  
 $\mu^{\zerogaupsi} \neq       \mu^{\widehat{\Psi}} $.

\subsection{Derivation of the Shape Jacobian}\label{derishjac}
We shall  now derive
formula \eqref{basicforJ}.  In order to do this, we shall compare the Laplace-Beltrami operator $\Delta_g$ on  absolute configuration space $\QA$  with the Laplace-Beltrami operator $\Delta_B$ on shape space $\QS=\QA/G$.
This comparison  would be easy  if we  could represent $\Delta_g$ in terms of coordinates $x^i =\{x^H, x^V\}$ such that the $x_V$ coordinate lines are all  inside the $G$-fibers  and the $x_H$ coordinate lines are orthogonal to them and thus  yield a  horizontal foliation. However, a coordinate system of this kind does not exist, not even locally, since
the existence of a  horizontal foliation of  $\QA$ is  precluded  by the curvature of the horizontal connection  arising from the rotations (see footnote \ref{nontrivial}). We shall not elaborate further on this.
Nonetheless a splitting into horizontal and vertical  components  can be obtained by expressing
 the Laplace-Beltrami operator
   in terms of a   basis  formed by a suitable set of horizontal and vertical vector fields,
   as  will be explained below.\footnote{Moreover,   the vertical vector fields that we shall need correspond to the Lie algebra of $G$, which is noncommutative, and thus do not arise from a coordinate system.}
 
 First, we  express gradient and divergence on a Riemannian manifold in terms of a general basis  $\X$ of vector fields  $ \X_i$, $ i=1, \ldots, n$. We replace  \eqref{gradcoo} by
 \begin{align}\label{gradgen}  \left( \grad_g\right)^i  = g^{ij} \X_j\,,
\end{align}
where $ g^{ij}= [g_{ij}]^{-1}$  with $g_{ij}= g(\X_i, \X_j)$,  and replace \eqref{divcoo}  with 
\begin{align}\label{divgen}  \div_g \Y =  \left( \frac{1}{\sqrt{|g|}} \X_i  \sqrt{|g|}  + \bomega^k \left([\X_k, \X_i]   \right)  \right) Y^i \,,\end{align}
where    $|g| = |\det(g_{ij})| $,  $\{\bomega^k\}$  is the dual basis in the cotangent space, i.e., $ \bomega^k (\X_i)= \delta_i^k$, $ [ \,\bigcdot\, ,\bigcdot\,]$ is the Lie bracket  (or commutator) of vector fields, and $Y^i$ are the components of $\Y$ with respect to the basis $\X_i$.\footnote{Formula \eqref{divgen}  is probably  in the literature, but  we have not succeeded in finding any reference. It is a straightforward consequence, for a manifold with a distinguished volume form (up to sign), of the fact that the divergence of a vector field times the volume form   is   the exterior derivative of the contraction of the vector field with the volume form.}  
Accordingly,   the Laplace-Beltrami operator \eqref{eq:LB}  becomes
\begin{align}\label{lapvecfi}
\Delta_g  &=\left( \frac{1}{\sqrt{|g|}} \X_i  \sqrt{|g|}  + \bomega^k \left([\X_k, \X_i]  \right) \right)g^{ij}\X_j \,,
\end{align}
which generalizes the standard formula  \eqref{eq:LB}  to an arbitrary basis.

Second, 
we  specify  a   basis   of vector fields     that is adapted to the geometrical structure  of  absolute configuration space  $\QA$  as a principal fiber bundle with base $\QS$ and fibers isomorphic to the similarity group $G$, in particular, to  the
orthogonal  decomposition of 
the  tangent space $T_\q \QA$ at any point $\q$ of   $\QA$ into    horizontal  subspace $T_\q \QA^H $ and vertical  subspace $T_\q \QA^V $ and the corresponding decomposition
$T  \QA= T  \QA^H \oplus T  \QA^V $ of  the tangent bundle.
In the   horizontal subspace we  choose as a   basis   the horizontal lift of a 
coordinate basis $X= \{X_\alpha\}$  in $\QS$,  $\alpha= 1, \ldots, n-7$. Note that  while a 
lift of the  vector field $X_\alpha$    is  not unique (as any vector field  on $\QA$ that projects down to  $X_\alpha$ represents a lift of $X_\alpha$),  there is only one horizontal lift  of $X_\alpha$  which we shall denote by
$ \widehat{X}_\alpha$. These  vector fields    form the  basis   $\X_H =  \{\widehat{X}_\alpha\}\equiv \widehat{X}$,  $\alpha= 1, \ldots, n-7$,  in the horizontal subspace.

In the vertical subspace we choose a  basis  formed by  vector fields 
that represent the action of the  infinitesimal generators of the group $G$ on  $\QA$. 
More precisely, observe that the action $\q\to \g(\q)$  of $G$ on $\QA$, $\g\in G$, defines,
for any given $\q\in\QA $ , the  map $\varphi_\q: G\to \QA$ given by  $\varphi_\q (\g)= \g(\q)$,
and the differential $\varphi_\q '$ of this map defines a map from $T_e(G)$, the tangent space to the identity $e$ of $G$, to $T_\q \QA$. Since  $T_e(G)$ is the   Lie-algebra  $\mathfrak{g}$ of the group $G$,  the image under  $\varphi_\q '$ of any  element  $\mathsf{L} $ of $\mathfrak{g}$ is a tangent vector at  $\q$ and, varying $\q$, one obtains the  vector field $\overline{\mathsf{L}}$  on $\QA$ associated with $\mathsf{L} $.  In particular, if  $\mathsf{L}_\beta$,  $\beta=1, \ldots 7$, are the generators of 
$\mathfrak{g}$,  their images   under  $\varphi_\q '$
form the  basis $\X_V= \{ \overline{\mathsf{L}}_\beta  \} $, $\beta=1, \ldots 7$, in the vertical  subspace. (It should be noted that
the vertical  vector fields so defined coincide with the image under $\varphi_\q '$ of the right invariant vector fields on $G$; in this regard, recall that the   Lie-algebra of the group can be equivalently defined  as the Lie-algebra  of the right ---or left --- invariant vector fields on $G$.)

Third, we  rewrite the  Laplace-Beltrami operator \eqref{lapvecfi} in terms of the basis $\X= \{\X_H, \X_V\}$ using the compact  (and slightly ambiguous) notation
\begin{align}
\Delta_g  &=\frac{1}{\sqrt{|g|}} \X_H  \sqrt{|g|}  g^{HH}\X_H +
\frac{1}{\sqrt{|g|}} \X_V  \sqrt{|g|}  g^{VV}\X_V\notag
\\  \label{lapvecfi2}
&\qquad+ \bomega^{A}   \left( [\X_A, \X_A] \right) g^{AA}\X  \,,
\end{align}
where repeated upper and lower indexes $H$, resp., $V$,  stands for summation over all
elements of   $ \{\X^H\}$, resp.,  $ \{\X^V\}$. In the last term the summation is over  $A=H,V$. Note, that  no mixed contributions $H$-$V$ occur, since  the vertical and horizontal vector fields are orthogonal and thus $g^{HV}=0$.  Consider  now 
$
\Delta_g \psibh$, the action of  $\Delta_g$ on   an  invariant function   $\psibh(\q) = \Psi(q) $.
  Since the second term in   \eqref{lapvecfi2} is  purely vertical, it gives no contribution. We rewrite the last term more explicitly, keeping only the non zero part of  its action on  invariant functions, to obtain
\begin{align}
\left( \bomega^H   \left([\X_H, \X_H] \right) +  \bomega^V   \left( [\X_V, \X_H] \right) \right) g^{HH} \X_H    \psibh \,. \end{align}
The first  term in the round brackets gives no contribution, in fact
\begin{align}
[\X_H, \X_H]  = [\widehat{X}, \widehat{X}] = \widehat{ [X, X]    } + \text{\sf Vertical}  =  \text{\sf Vertical}\,,
\end{align}
where in the last equality we have used the fact that  $X$ is a coordinate basis, and thus its elements commute; moreover,  $ \bomega^H ( \text{\sf Vertical}
 ) $  is clearly zero. As for the second term, expressing the commutator of the vector fields by means 
 of the Lie derivative $ \mathcal{L}$,
 \[
  [\X_V, \X_H] = \mathcal{L}_{\X_V} \X_H = 0
 \]
by symmetry, i.e., the $G$-invariance of the vector fields $\X_H $.
We conclude that for $\psibh $ an invariant function on  $\QA$, we obtain
\begin{align}\label{deltagoninv}
\Delta_g\psibh  
=\frac{1}{\sqrt{|g|}} \X_H  \sqrt{|g|}  g^{HH}\X_H \psibh\,.
\end{align}

Fourth, we consider the  Laplace-Beltrami operator on shape space acting on $\psi = \psi(q)$,
\begin{align}\label{deltabb}
\Delta_B \psi   =\frac{1}{\sqrt{|g_B|}} X  \sqrt{|g_B|}  g^{BB}  X \psi \,.
\end{align}
Then the action of a lift of $\Delta_B$ on the  invariant function   $\psibh = \psibh(\q)$ associated with $\psi$  is given by
\begin{align}\label{deltabh}
 \lDeltag  \psibh  
=\frac{1}{\sqrt{|g_B|}} \X_H  \sqrt{|g_B|}  g^{HH}  \X_H  \psibh\,.
\end{align}
Comparing \eqref{deltabh} with  \eqref{deltagoninv},   
we write this as
\begin{align}\label{deltasb}
\lDeltag   \psibh =   \jac  \frac{1}{\sqrt{|g|}} \, \X_H   \jac^{-1} \sqrt{|g|} g^{HH}  \X_H  \psibh  
\end{align}
with
\begin{align}\label{forshjac}
\jac = \frac{\sqrt{|g|}}{ \sqrt{|g_B|} } \,.
\end{align}

Fifth (and finally), we consider the operator 
\[ O = \jac  \, \div_g \jac^{-1} \grad_g \,,\]
with $\jac$ given by \eqref{forshjac},
and observe that
\begin{multline} \jac  \, \div_g \jac^{-1} \Y   =  \jac   \left( \frac{1}{\sqrt{|g|}} \X_i  \sqrt{|g|}  + \bomega^k \left([\X_k, \X_i]   \right)     \right) \jac^{-1}  Y^i \\
 = \left( \frac{\jac}{\sqrt{|g|}} \X_i  \jac^{-1} \sqrt{|g|}  + \bomega^k \left([\X_k, \X_i]   \right)     \right)    Y^i \,.
\end{multline}
Thus $O\psibh$, with $\psibh$ an  invariant function,  coincides with the right hand side  of 
\eqref{deltasb} (for the same reasons that led us from \eqref{lapvecfi2} to \eqref{deltagoninv}).  Therefore  $O$, i.e.   \eqref{basicforJ}, is a lift of $\Delta_B$, with  
the  shape Jacobian $\jac$  given explicitly by equation~\eqref{forshjac}.
 
\subsection{Computation of the Shape Jacobian}\label{compshjac}
Our last task is to  derive formula \eqref{forforJ}  from equation  \eqref{forshjac} for   the  shape Jacobian $\jac$.

First,  we observe that in the basis $\{\X_H, \X_V\}$ the metric $g$ has the block diagonal decomposition (slightly abusing notation)
\[
g =  \begin{pmatrix}  g_V & 0\\ 0 & g_H    \end{pmatrix}\,,
\]
where $g_H$   can be identified with  $ g_B$  and $g_V$ is  the restriction of  $g$ to the vertical vector fields. Since $|g| =
|g_V| |g_H|$ and $|g_B|= |g_H|$, it follows from \eqref{forshjac}  that
\begin{align}\label{J}
\jac = \frac{\sqrt{|g|}}{ \sqrt{|g_B|} } =   \frac{\sqrt{ |g_V| |g_H|}}{ \sqrt{|g_B|} } = \sqrt{|g_V|} \,,
\end{align}
so that $\jac$ turns out to be the invariant volume density in the vertical subspace with respect to fiber volume element corresponding to $\X_V$.   Moreover, by \eqref{eq:bmetrxxx1}  we have that
\begin{align}\label{shja1}
\jac = f^{7/2} \jac_e\,,
\end{align}
where $\jac_e$ is the vertical volume  density for the  mass-weighted Euclidean metric \eqref{eq:euclid} instead of the invariant metric $g$ involving the conformal factor $f$.  For this we have 
\begin{align}\label{shja2}
 \jac_e (\q) = \vol (\X_V) (\q)\,,
\end{align}
the $7$-dimensional (Euclidean) volume of the parallelepiped  in $T_\q \QA^V $ generated by $\X_V$, i.e., by the vertical tangent vectors at $\q$ obtained by evaluating at $\q$ the $7$  vector fields that generate $G$.

Second, we may split 
\begin{align}  \X_V = (\X_\text{tr}, \X_\text{rs})\,,
\end{align}
where $\X_\text{tr}$ refers to the $3$ generators of translations and  $\X_\text{rs}$ to the $4$ generators of rotations and scaling. However, the vector fields $\X_\text{rs}$ are not in general orthogonal to those of $\X_\text{tr}$. We therefore consider also 
$\widetilde{\X}_\text{rs} = P_\text{tr}^\perp \X_\text{rs} $, the orthogonal projection of the vectors $\X_\text{rs}$ into the orthogonal complement of the subspace of the tangent space corresponding to translations. We thus have that
\begin{align}
 \vol (\X_V) &=\vol (\X_\text{tr}) \cdot \vol (\widetilde{\X}_\text{rs}) \\
 & \equiv   \vol (\X_\text{tr}) \widetilde{\jac}_\text{rs}  = \const \widetilde{\jac}_\text{rs}\,,
\end{align}
since $\vol (\X_\text{tr})$ is a constant, independent of $\q$ (which we may take to be $1$ by letting the translation vectors in $\X_\text{tr}$ to be orthonormal). It should be observed that 
$\widetilde{\X}_\text{rs}$ consists of the generators of rotations and scalings  about the center of mass of the configuration $\q$. To see this, note that if we represent ${\q}$ in center of mass and relative coordinates  $\ovqc =  \q -  \vect{q}_\text{cm}$, i.e.,   $\q= (\vect{q}_\text{cm}, \ovqc) $, then for any rotation or scaling $\g$, we have that the action of $\g$ on $\q$ is given in these coordinates by  $\g (\vect{q}_\text{cm}, \ovqc) =  (\g(\vect{q}_\text{cm}),  \g(\ovqc)) $, while for the corresponding action $\widetilde{\g}$ about the center of mass, 
$\widetilde{\g} (\vect{q}_\text{cm}, \ovqc) =  ( \vect{q}_\text{cm},  \g(\ovqc)) $.

Third,  we may split 
\begin{align} \widetilde{\X}_\text{rs} = (\widetilde{\X}_\text{rot},\widetilde{\X}_\text{s})\,
\end{align}
into generators $\widetilde{\X}_\text{rot}$ of rotations about the center of mass and a generator $\widetilde{\X}_\text{s}$ of scalings about the center of mass. Then we have that
\begin{align}\label{prodovolel3}
\widetilde{\jac}_\text{rs} =   \vol (\widetilde{\X}_\text{rs}) =  
\vol (\widetilde{\X}_\text{rot}) \cdot \vol (\widetilde{\X}_\text{s})\,,
\end{align}
since  $\widetilde{\X}_\text{s}$ is orthogonal to $\widetilde{\X}_\text{rot}$.

Fourth, 
we have that
\begin{align} \label{forforr}
\vol (\widetilde{\X}_\text{s}) = \rr
\end{align} 
up to a constant, independent of $\q$, where $\rr=\rr(\q)$ 
is given by  equation~\eqref {mominsca}. To see this, note that the effect of a scaling at $\q$ about the center of mass is proportional to the value of $\rr$ at $\q$. As for the other volume element in  \eqref{prodovolel3}, we have
\begin{align} \label{widehatxrot}
\vol (\widetilde{\X}_\text{rot}) =  \sqrt{\det \mathsf{M}}\,,
\end{align} 
where $\mathsf{M}$ is    the tensor  of inertia of the configuration $\q$  about the center
of mass, whose matrix elements with respect to an orthogonal cartesian system $x$,$y$,$z$ 
are given by equation \eqref{mateltensin}.
This formula for the volume element is presumably standard.
A way to see   how it comes about is the following.

To simplify the notations, let us drop ``tildas'' and ``rot'' and stipulate that in this paragraph  $ (\X_i)$, $i=x,y,z$, denotes a basis for  the generators of rotations about the center of mass
$\vect{q}_\text{cm} $ ($x,y, z$  refer to any orthogonal frame  with origin in the center of mass). 
Then the volume element in \eqref{widehatxrot} is given by 
$\sqrt{\det \mathsf{A}}$, 
where $\mathsf{A}$ is the matrix with entries
$A_{ij} = g_e\left(\X_i, \X_j    \right)$.
For $ \ovqc =  \q -  \vect{q}_\text{cm} $   a configuration relative to the center of mass,   
let $\ovqc= (\ovq_1,\ldots , \ovq_N )$.
Observe  that  a   generator  corresponding to the action of  a rotation on configurations   is of the form $ \X_\Om  ( \ovqc ) = ( \Om  \times  \ovq _1, \ldots, 
\Om  \times  \ovq _N)$, where  $\Om$ is a 3-dimensional vector of components $\Omega^i$ (with respect to the $xyz$ frame).
%
%
%
Thus the 3-dimensional Lie algebra corresponds naturally to the 3-dimensional vectors $\Om$, with $\Omega^i $ being the coordinates of a general element $ \X_\Om $ of the Lie algebra  with respect to the basis  $ (\X_i)$. A vector $\Om$ 
corresponds to a general instantaneous rotational motion. Consider the kinetic energy
\begin{align} \label{kinenerrot} K=  \frac12 g_e (\X_\Om , \X_\Om)    \end{align}
for such a motion. On the one hand, it is known to be given by 
\begin{align} \label{kinenerrot2}  K = \frac12 M_{ij}   \Omega^i \Omega^j \,,
  \end{align}
where  $ \mathsf{M}= \{M_{ij} \} $ is  the moment of inertia tensor. On the other hand, expanding  the right hand side  of \eqref{kinenerrot} by expressing  
$
\X_\Om = \sum_i \Omega^i \X_i 
$  in the basis $ (\X_i)$,    
one obtains 
\begin{align*}
g_e (\X_\Om , \X_\Om) =   g_e\left(\X_i, \X_j    \right) \Omega^i\Omega^j = A_{ij} \Omega^i\Omega^j \,.
\end{align*}
 Thus  equating the right hand sides of \eqref{kinenerrot} and \eqref{kinenerrot2}, one sees that the matrix $\mathsf{A}$ is indeed the tensor of inertia $\mathsf{M}$, whence equation \eqref{widehatxrot}.

Fifth (and finally),  substituting in \eqref{shja1} the formula for $\jac_e$ given by
\eqref{shja2}, with $ \vol (\X_V)= \widetilde{\jac}_\text{rs}$, and using formulas
\eqref{prodovolel3},\eqref{forforr},  and \eqref{widehatxrot} for  
$ \widetilde{\jac}_\text{rs}$, we have that
\[ 
\jac= \rr f^{7/2}  \sqrt{\det \mathsf{M}}\,, \] 
which is  formula \eqref{forforJ} for the shape Jacobian.

\subsection{More Gauge Freedom}

As we have already stressed, the  lift
 to $\QA$  of the Laplace-Beltrami operator $\Delta_B$ on  $\QS$ is by no means unique.  The  
 ``canonical lift''  \eqref{basicforJ}, with $\jac$ given by equation   \eqref{forforJ}, is very natural, but other choices are possible.  This lack of uniqueness increases the gauge freedom we have in defining the 
{Schr\"odinger gauge}.
In particular, we may use this freedom  
 to define a shape Jacobian that is an invariant function  on  $\QA$, i.e., a function of the configuration $\q$ which depends only on its shape.


Note that $\jac$ is not invariant: $f^{7/2}$ scales like $\rr^{-7}$ and   $\sqrt{\det \mathsf{M}} $ like $\rr^3$. We thus have that 
\begin{align} \jac = \rr^{-3} \jac_B \,,\end{align}
where \begin{align} \jac_B = f_1^{7/2}   \sqrt{\det \mathsf{M}_1} \end{align}
is invariant. Here the subscript $1$ indicates the quantities have to be evaluated, not at $\q$, but at $\q_1$, the configuration with $\rr=1$ obtained by rescaling $\q$.

To define a lift  $\lDeltag$ of $\Delta_B$, we could as well have used the invariant 
\begin{align}\label{formulaJB}
\jac_B = \rr^3 \jac =\rr^4  f^{7/2}  \sqrt{\det \mathsf{M}} 
\end{align}
 instead of $\jac$. This would have in no way affected the results and the arguments in Sect.
\ref{sec:qmsgpr} and  Sect. \ref{derishjac},  though it would yield somewhat  different potentials $V$ and $U$ in \eqref{eq:Steptwo} and \eqref{step312}, respectively.

\subsection{The Canonical Conformal Factor}\label{The Canonical Conformal Factor}
Instead of computing $\jac_B$ for a  given $f$, we might  read \eqref{formulaJB} the other way round, and ask what is the conformal factor that gives rise to the simplest $\jac_B$.
The simplest possibility is $\jac_B= 1$  and this is associated 
with $f(\q) \equiv f_c (\q)$ given by equation \eqref{pconfacty}, i.e., 
\[
f(\q) =f_c(\q)\equiv   \rr^{-\frac{8}{7}} (\det\mathsf{M})^{-\frac{1}{7}}\,,
\]
 the {\em canonical conformal factor}. Note that replacing $\jac$ in   \eqref{V_1}     with 
 $\jac_B= 1$ gives  $V_1=0$ so that the potential  in  the Hamiltonian  \eqref{eq:Steptwo} is $V=V_2$, with the form   \eqref {exvtwo} of  $V_2$  unaffected.
(Of course, one needs to evaluate  it   for $f= f_c$.) Similarly, the potential  
$U$ in \eqref{step32} becomes
\begin{align}
\label{step32cc}
U= -f_c \mathscr{E}   - \frac{\hbar^2}{8} \frac{n-2}{n-1} f_c  R_{g_c}\,,
\end{align}
where $R_{g_c}$ (cf. \eqref{curpot}) is now  the scalar curvature of the metric $g=g_c$ associated with $f_c$.

\section{Subsystems}\label{ss}
\subsection{ Conditional Wave Functions}

In physics we are usually concerned not with the entire universe but with subsystems of the universe, for example with a hydrogen atom or a pair of entangled photons. The quantum mechanical treatment of such systems  involves the quantum state of that system, often given by its wave function---not the wave function of the universe. Bohmian mechanics provides a precise formulation and understanding of this notion in terms of the conditional wave function \cite{durr1992quantum}
\begin{equation}\label{conprobfor}
\psi(x)=\Psi(x,Y)\,,
\end{equation}
where $\Psi=\Psi(q)=\Psi(x,y)$ is the wave function of the universe, 
with $x$ and $ y$   the generic variables for the configurations of the system and its environment, respectively, and  where $Y$ is  the actual configuration of the environment. The conditional wave function of a Bohmian system behaves exactly as one would expect the wave function of a system to behave, with respect to both  dynamics and statistics. It is natural to ask how and whether  the conditional wave function can be defined for Bohmian mechanics on shape space.

For this the following problem arises. There is no natural product structure
$$\QS=\QS_{sys} \times \QS_{env}$$ for shape space: Here  the system is a collection of (labelled) particles with its own shape space $\QS_{sys}=\mathcal{X}$, the set of possible shapes $X$ of the system, and the environment consists of the rest of the particles of the universe, with shape space $\QS_{env}=\mathcal{Y}=\{Q_{env}=Y\}$, with $Y$ the shape associated with the particles (labels) of the environment.
The crucial fact is that it is not true that 
$$\QS=\mathcal{X} \times \mathcal{Y}.$$
$X$ and $Y$ don't involve sufficient information to determine the complete shape $Q$. What is missing is the spatial relationship between these shapes. 

Nonetheless we have that $\QS$ can be identified with
$$\mathcal{X}_Y\times_y\mathcal{Y}=\{(X,Y)|Y\in\QS_{env}, X\in\QS_Y\}\,,$$
where
$$\QS_Y=\{Q\in\QS| Q_{env}=Y\}.$$

We may then define the conditional wave function for the subsystem, for $Y\in\QS_{env}$ and universal wave function  $\Psi$, by
\begin{equation}\label{cwfu}\psi(x)=\Psi(x,Y), \quad x\in \QS_Y. \end{equation}
This looks like the usual conditional wave function, but it is important to bear in mind that, unlike with the usual conditional wave function, here $x$ represents the shape of the universe for a fixed $Y$ and there are no obvious natural coordinates to efficiently describe it. 

To obtain such a thing the notion of a frame might be useful: Given $Y\in\QS_{env}$, a frame $\lift{Y}$ is a choice of point in the fiber over $Y$ in the absolute configuration space of the environment. ($\lift{Y}$ corresponds roughly to the usual notion of frame of reference.)

Given $Y$  and a frame $\lift{Y}$, we obtain natural coordinates for $\QS_Y$: Given $X\in\QS_Y$ there is a unique $\lift{X}\in\mathbb{R}^{3M}$ (with $M$ the number of particles of the subsystem, which  we shall assume   from here on is   such that  the number of particles  in the environment $N-M\ge 3$) such that $(\lift{X},\lift{Y})$ is in the fiber above $X\in\QS_Y$.  The map $X\mapsto\lift{X}$ is a one-to-one correspondence between $\QS_Y$ and $\mathbb{R}^{3M}$.

Given the frame $\lift{Y}$, we may represent the conditional wave function by
\begin{equation}\label{psihat} \liftw{\psi}(\lift{x})=\psi(x), \end{equation}
where $x$ corresponds to $\lift{x}$. In other words,  
\begin{equation}\label{psihat2} \liftw{\psi}(\lift{x})=\liftw{\Psi}(\lift{x}, \lift{Y}), \end{equation}
with $\liftw{\Psi} $ the lift of $\Psi$ to absolute configuration space,  $\lift{Y}$  a lift of $Y$, $\lift{x}$ the lift of $x$ in the frame $\lift{Y}$, and the ``hat'' refers to any of the  gauge  equivalent representations  $\firstgaupsi$, 
 $\secondgaupsi$,
  $\thirdgaupsi$
(and presumably $\zerogaupsi$) of the universal wave function    that  we have  described in Sect. \ref{sec:qmsgpr}.  
Thus, the  absolute configuration space level  conditional wave functions 
$\gauconpsi_1$, $\gauconpsi_2$ and $\gauconpsi_3$ (and $\gauconpsi_S$) are  at the same time  different representations of the shape space conditional wave function.
Moreover, the $\liftw{\psi}$s obtained using different frames are in an appropriate sense equivalent.

This conditional wave function $\liftw{\psi}$ behaves like the wave function of the subsystem, both with respect to the dynamics of configurations, via the guiding equation, and with respect to probabilities  for the subsystem, via what has been called  the fundamental conditional probability formula \cite{durr1992quantum}. With regard to the dynamics, this is clear from the form of the dynamics on absolute configuration space. The latter, while true, is not at all so clear. That it is so follows from the analysis in   Sect. \ref{probonpaths}.
%


\subsection{Subsystems and the Role of Projectivity}
The time-parameter corresponding to the use of the denominator in footnote \ref{fn:proj}  has the nice feature that the dynamics using that time-parameter depends on fewer details of the wave function than would be the case if the denominator were deleted: with the denominator the dynamics depends only on the ray of $\psi$, with $\psi$ and $c\psi$ yielding the same dynamics for any constant $c\neq0$. This has a particularly nice implication for the behavior of subsystems. With this choice of time-parameter the dynamics for a subsystem will often not depend upon the configuration of its environment, with the subsystem evolving according to an autonomous  evolution involving only  the configuration and the (conditional) wave function of the subsystem itself \cite{durr1992quantum}.
This would happen when the subsystem is suitably decoupled from its environment, for example for a product wave function when there is no interaction between system and environment. Without the denominator this would not be true, and there would appear to be an additional  nonlocal dependence of the behavior of a subsystem on that of its environment that would not be present with a time-parameter associated with the use of the usual denominator.
%
\subsection{The Emergence of Metrical Time}

By metrical time we refer to any objective physical coordination of the configurations along a  geometrical path  in a configuration space with the points of a one-dimensional continuum: a (continuous) mapping from the continuum onto the path. The continuum is usually represented by the real numbers, but it need not be. However, it should be physically distinct from the particular continuum that is the path itself.

Understood in this way, metrical time does not exist, from the relational point of view, for the universe as a whole. However, for subsystems of the universe metrical time naturally emerges: the continuum with which the geometrical path corresponding to the evolution of the subsystem is coordinated can be taken to be the path of its environment, with the obvious mapping between the paths.

\section{Probability }\label{probonpaths}
{

\subsection{Bohmian Mechanics and Probability}\label{bmap}
In Bohmian mechanics, for a non-relativistic system of particles, the configuration of a system is regarded as random, with randomness corresponding to the
 quantum equilibrium distribution $\mu^\psi$ given by $|\psi|^2dq$. What this actually means, in a deterministic theory such as Bohmian mechanics, is a delicate matter, involving a long story \cite{durr1992quantum} with details and distinctions that we shall ignore here. However a crucial ingredient for that analysis---for an understanding of the origin of quantum randomness in a universe governed by Bohmian mechanics---is the {\it fundamental conditional probability formula} for the conditional distribution of the configuration $X_t$ of a system at time $t$ given that of its environment~$Y_t$ at that time:
\begin{equation}\label{condprob}
 P^{\Psi_0}(X_t\in dx\, | Y_t) = |\psi_t(x)|^2 dx, 
\end{equation}
 where $\Psi_0$ is the initial wave function of the universe and $P^{\Psi_0}$ is the probability distribution on trajectories arising from the Bohmian dynamics with an initial quantum equilibrium distribution, and $\psi_t$ is the (normalized) conditional wave function \eqref{conprobfor} of the system at time $t$.
 
 A crucial ingredient  in proving \eqref{condprob} is {\em equivariance} \cite{durr1992quantum}:  if  at any time   the system  configuration is randomly distributed 
 according to $|\Psi_0|^2$, then at any other time $t$ it will be distributed according to 
 $|\Psi_t|^2$. Equivariance  is an immediate consequence of the continuity equation  arising from the Schr\"odinger equation:
 \begin{equation}  \label{continuityequation}
 \frac{\partial \rho^\Psi}{\partial t} + \div J^\Psi = 0\,,
 \end{equation}
with   $\rho^\Psi= |\Psi|^2$, the quantum equilibrium distribution,  and 
\begin{equation} \label{quantumcurrent}
J^\Psi = \rho^\Psi v^\Psi \,,
\end{equation}
 the quantum probability current,  where $v^\Psi$ is  the Bohmian velocity
in the right hand side of \eqref{gl},
\begin{equation} \label{bohmvelocity}
v^\Psi=  \hbar  \, \mathrm{ Im } \frac{ \nabla   {\Psi}}{ {\Psi}} 
\end{equation}
(as usual here, the masses  are absorbed in the gradient).

 More generally, for Bohmian mechanics on a Riemannian manifold with metric $g$, the quantum equilibrium distribution $\mu^\Psi$ is given by $|\Psi|^2d\mu_g$. In terms of any coordinate system $x=(x_1,\ldots,x_n)$, we have that \begin{equation}\label{dmugdx} d\mu_g=\sqrt{|g|}dx,\end{equation} where $dx=dx_1\cdots dx_n$ and $g=g_{ij}=g(\X_i,\X_j)$, with $ \X_i =\partial/\partial x_i$.

Moreover, as already indicated several times,   the Bohmian velocity on a Riemannian manifold  
is still given by  \eqref{bohmvelocity} with $\nabla= \nabla_g$,  the gradient with respect to the metric $g$. Furthermore,   letting  $\rho^\Psi $   be the density of 
$\mu^\Psi$ with respect to  $\mu_g$,  i.e.,  $\rho^\Psi = |\Psi|^2$,    the current is  most conveniently  represented 
as a density with respect to  $\mu_g$   as in 
 \eqref{quantumcurrent};  in this representation
 the continuity equation \eqref{continuityequation} holds for $\div=\div_g$, the divergence with respect to $g$. So, all the Bohmian structure, including the quantum equilibrium measure,  transfers straightforwardly from the familar Euclidean setting  to a general Riemannian manifold.

%
%
%

\subsection{The Problem of Non-Normalizable Measures}\label{pnnm}
Note that since it is translation and scaling invariant, the wave function in the \Sc gauge or in any of the  gauges discussed in Sect.\ref{scg}) are not normalizable.
} In other words, 
 $\mu^{\zerogaupsi} $  and  $ \mu^{\widehat{\Psi}} $
 given by \eqref{muphi} and  \eqref{born1}, respectively, 
are   non-normalizable. 
%
%
%
%
%
{
However, since the non-normalizability arises from 
unobservable (and, from a shape space point of view, unphysical) 
differences and dimensions it should somehow not be a problem. 
%

Nonetheless, the real question is how  the empirical 
distributions arising from the fundamental shape space level compare with those 
coming from the physics in a gauge. While the different gauges, such as the \Sc 
gauge, correspond to theories that, we argued,  are empirically equivalent to 
the fundamental shape space theory,  that was only in purely dynamical terms. 
We have not yet addressed the possible differences in empirical distributions 
that may arise. We would like to see that they don't. 

There are several considerations that suggest that the non-normalizability should not be a genuine problem:
\begin{itemize}
\item[1)]As just mentioned, the non-normalizability arises only from non-observable dimensions, suggesting that it should be physically  irrelevant.
\item[2)] It is the universal wave function $\gaupsi$ (in any of the   gauges) that is not normalizable. But the  universal wave function is rarely used in practice. In quantum  mechanics we usually deal, not with the entire universe, but with small subsystems of the universe. The wave functions with which we usually deal are thus conditional wave functions, and there seems to be no reason why these should fail to be normalizable.
\item[3)]  In statistical mechanics the Lebesgue  measure on the phase space for a gas in a box is non-normalizable. Why is this not a problem? It is because the energy is a constant of the motion, and the restriction of the Lebesgue measure to the energy surface is normalizable (for suitable potentials). Similarly here, with Bohmian mechanics in the \Sc gauge, or  in any of the  gauges on absolute configuration space, the center of mass and the moment of inertia about the origin, or about the center of mass, are constants of the motion. Thus,   it would seem that the appropriate  measure that we should be considering here is $
\mu^{\gaupsi}_{\Gamma}$, the conditional distribution for $\mu^{\gaupsi}$ given $\Gamma$, the one given by $|\gaupsi|^2$ on a surface $\Gamma$ of constant center of mass and  moment of inertia about the origin, and not on the entire absolute configuration space, and this is presumably normalizable. Moreover, all such probability distributions, for different choices of $\Gamma$, are physically equivalent, since they correspond to the same probability distribution on shape space. 
\end{itemize}
 
The probability distributions  described in 3), while they seem to correspond to the appropriate measures  on shape space, appear to be entirely inappropriate insofar as the fundamental conditional probability formula is concerned. For example, for a single-particle system the configuration of that system  would be completely determined  by  the configuration of its environment,  rather than being randomly distributed according to the quantum equilibrium distribution. What gives?

\subsection{The Fundamental Conditional Probability Formula for Evolving Wave Functions}\label{fundaconform}
In order to answer the last question, as well as to obtain a sharp resolution of the non-normalizability issue and an understanding of how to carry out the usual quantum equilibrium analysis \cite{durr1992quantum} yielding the Born rule, the following observation is crucial: The conditional distribution, arising from $\mu^{\gaupsi}$, 
of the configuration of a system given its environment in absolute configuration space is unphysical, and is thus not relevant to an appropriate quantum equilibrium analysis. That is because (from the shape space point of view) the absolute configuration of the environment  is unphysical. What is physical, and what we should be conditioning on, is the shape of the environment. And when we condition on this, the result will be given in terms of the conditional wave function $\gauconpsi$~\eqref{psihat}.

In fact, on the fundamental level, with respect to the quantum equilibrium distribution  on shape space, $ d \mu^{\Psi}_{B} = |\Psi |^2 d\mu_B $ (where $d\mu_B  \equiv d\mu_{g_{B}}$), the conditional distribution  of the configuration of a system (i.e., of the configuration of the universe, see \eqref{cwfu}) given the shape of its environment
is perhaps most naturally expressed on the absolute level, via the use of a frame and of the corresponding conditional wave function  $\gauconpsi$ in any one of the first three gauges---for example,    the  conditional wave function in the first gauge we considered, namely the straightforward  lift   \eqref{strhl}---using the frame $\lift Y$ rather than the shape $Y$. And since the quantum equilibrium distribution for this conditional wave function represents the corresponding conditional distribution on shape space {(as we shall argue below)}, it follows that the conditional wave function on absolute configuration space is normalizable.

 That this is in fact so, i.e.,  that the lift to absolute configuration space of the fundamental conditional probability formula on shape space is given by the Born rule for the conditional wave function on absolute configuration space, can be seen as follows:

\begin{itemize}
\item Unlike marginal distributions,  conditional measures are well defined, up to a constant multiple, even for a non-normalizable measure $\mu$.

\item Suppose we  condition on something, for example, the environment $\lift Y$ of a subsystem, corresponding to  a  leaf $\Gamma_\mathscr{E}$ of a foliation $\mathscr{E}$ (e.g., into the level sets of the random variable $\lift Y$). We would obtain the same result if we  had first conditioned on  a leaf  $\Gamma_\mathscr{F} $,   $\Gamma_\mathscr{E} \subset \Gamma_\mathscr{F} $, of a  coarser  foliation\footnote{identifying $\mathscr{F}$ and $\mathscr{E}$  with the $\sigma$-algebras they generate, consisting of the 
measurable sets that are unions of the leaves of the 
respective foliations}   $\mathscr{F}\subset \mathscr{E}$,  obtaining the conditional measure 
$\mu^{}_{\Gamma_{\!\mathscr{F}}}$,
and then, with respect to  $\mu^{}_{\Gamma_{\!\mathscr{F}}}$,
conditioned on  $\Gamma_\mathscr{E}$.
\item For $\mu= \mu^{\gaupsi}$, the (non-normalizable) quantum equilibrium distribution on absolute configuration space, we can choose $\mathscr{F}$ so that the measures  $\mu^{}_{\Gamma_{\!\mathscr{F}}}$ all correspond to the quantum equilibrium distribution on shape space, and $\mathscr{E}$ so that it corresponds to the shape of the environment.
\item If this can indeed be done we obtain our desired result.
\end{itemize}

In more detail,  fix  now the subsystem, and let $\mathscr{E}$ correspond to the configuration $\lift Y$ of its environment.  $\mathscr{F}$ must be chosen so that the following is true: Each leaf of $\mathscr{F}$ must provide a representation of shape space as a measure space. Not only must there be a smooth bijection between shape space and each leaf of $\mathscr{F}$, but under this bijection  we must have that each of the measures  $\mu^{}_{\Gamma_{\!\mathscr{F}}}$
 corresponds to $\mu^{\Psi}_{B}$. Moreover, we must also have that (i) the leaf of $\mathscr{F}$ to which an absolute configuration belongs is determined by the configuration  of its environment, i.e.,  $\mathscr{F} \subset { \mathscr{E}}$,   and that (ii) different leaves  of $\mathscr{E}$ belonging  to the same leaf of $\mathscr{F}$  correspond to  environmental configurations with different shapes, so that 
with respect to the bijection, the configuration of the environment corresponds to its shape.

Such an $\mathscr{F}$ can be generated from a cross-section of the absolute configuration space of the environment regarded as a bundle over its shape space. Such a cross-section naturally induces  a cross-section in the universal absolute configuration space regarded as a bundle over shape space (since for any shape there is a unique absolute configuration compatible with the environmental cross-section). The cross-section so obtained provides a single leaf $\Sigma_1$ of $\mathscr{F}$; the other leaves of $\mathscr{F}$ are obtained by the application of the symmetry group $G$ to $\Sigma_1$. In this way, absolute configuration space can be identified with $G\times\Sigma_1$, with $\mathscr{F}$ corresponding to $G$, i.e., having leaves $\Sigma_{\g}=\g\Sigma_1$.

Note that with this  $\mathscr{F}$ we have, essentially by construction, that the shape of the environment corresponds,  on each leaf of $\mathscr{F}$, to its absolute configuration. Note also that this would not be true for the foliation corresponding to the (quotient under rotations of the) surfaces $\Gamma$  described  in 3)   of Sect. \ref{pnnm}, for which a single shape of the environment would correspond to many different absolute environmental configurations on each leaf (even after rotations have been factored out).

We now check that for this \F the conditional measures $\mu^{}_{\Gamma_{\!\mathscr{F}}}$ correspond to  the quantum equilibrium distribution on shape space. For this we will use the following general formula for the Riemannian volume element $\mu_g$ in terms of a general basis of vector fields $\X_i$:
$$ 
d\mu_g= \sqrt{|g|}\,d\omega_1\cdots d\omega_n\,,
$$
where $|g|$ is defined below \eqref{divgen}. This formula is similar to \eqref{dmugdx}  for the case when the vector fields are coordinate  vector fields, but with $dx_1\dots dx_n$ replaced by $d\omega_1\cdots d\omega_n$, the volume element arising from the $n$-form $d\omega_1\wedge\dots \wedge d\omega_n,$ where $\omega_1,\dots, \omega_n$ is the basis of 1-forms dual  to the basis $\X_1,\dots, \X_n$ of vector fields, see Sect. \ref{derishjac}.

For the basis of vector fields $\X=(\X_V,\X_H)$ described there, we obtain that
$$
d\mu_g= \sqrt{|g|}\ d\omega_V d\omega_H
=\sqrt{|g_V|}\sqrt{|g_H|}\ d\omega_V d\omega_H,
$$
where $\ d\omega_V$ is the volume element arising from $\X_V$, a volume element on the fibers of absolute configuration space, and $d\omega_H$ corresponds to the volume element on shape space arising from the coordinate system involved in the definition of $\X_H$. 

Since $\sqrt{|g_H|}\ d\omega_H$ corresponds to $\sqrt{|g_B|}dx=d\mu_B$, where $dx$ is the coordinate measure on shape space for these coordinates, we have that $d\mu_g$ corresponds to $\sqrt{|g_V|}\ d\omega_V d\mu_B.$ Moreover, $d\omega_V$ is, up to a constant factor, the image of the right Haar measure $\mu_G$ on $G$. We thus have, using the representation $G\times\Sigma_1$ for absolute configuration space, that
$$
d\mu_g= \jac d\mu_Gd\mu^1_B,
$$
where  {$\jac$ is given in \eqref{J} and} $\mu^1_B$ is the image of $\mu_B$ on $\Sigma_1$.

Now since $\widehat{\Delta}_B$, see  Sect. \ref{scg}, is self-adjoint, not with respect to $d\mu_g$,  but with respect to  $\jac^{-1}d\mu_g= d\mu_Gd\mu^1_B,$ we have that, for $\widehat{\Delta}_B$, and any wave function $\Psi$  on shape space, the quantum equilibrium distribution $\mu^\Psi$ is given by
$$
d\mu^{\gaupsi} = |\widehat{\Psi}|^2 d\mu_Gd\mu^1_B.
$$ 
Thus, for $\mu^{\gaupsi}$, the conditional probability distribution given $G$ is  $|\widehat{\Psi}|^2 d\mu^1_B,$ i.e., the image of $\mu^{\Psi}_B$, just as we wanted.

Finally,   as we   indicated earlier in Sect. \ref{rmkbm}, as we proceed through the first three gauges,  each involving its own wave function and measure for self-adjointness,  the transformations connecting the gauges have been so defined as to leave invariant the corresponding quantum equilibrium distributions $\mu^{\gaupsi}$. For each gauge, we are in fact dealing with the same measure on absolute configuration space, and hence the same conditional measure given  a leaf $\Gamma_{\mathscr{F}}$ of the foliation \F and the same conditional measure given a leaf  $\Gamma_{\mathscr{E}}$  of the foliation \E\!. Thus in each gauge, the conditional  distribution of the quantum equilibrium measure given the configuration of the environment yields, in fact, the conditional distribution on shape space given the shape of the environment, which is what we wanted to establish.

In summary,  we have  obtained that  the conditional distribution of the configuration $X$  of a system given the shape  $Y$ of its environment  is given by
\begin{align}\label{last1} \mu_B^\Psi (X\in dx\, | Y)  =  {\mu}^{\widehat{ \Psi}} (\lift{X}\in d\lift{x} | \lift{Y})\,.
\end{align}
As a consequence, we have (from equivariance)
 the {\em fundamental conditional probability formula on shape space}, for the conditional distribution of the configuration $X_t$  of a system at time $t$ given the shape  $Y_t$ of its environment at that time.
 
 \medskip 
\noindent{\bf Fundamental conditional probability formula on shape space:} {\em 
 \begin{align}
P^{\Psi_0}(X_t\in dx\, | Y_t= Y) =   {\mu}^{\widehat{ \Psi}_t} (\lift{X}\in d\lift{x} | \lift{Y})\,,
\end{align}
where $\Psi_0$ is the initial  universal wave function on shape space
(at $t=0$), $ \lift{Y}$ is a lift of $Y$, $\lift{x}$ the  lift of $x$ in the frame $\lift{Y}$,   
 $\Psi_t$ is the universal wave function at time $t$, with $\mu^{\gaupsi_t} $   the non-normalizable quantum equilibrium measure on absolute configuration space---the  lift
\eqref{born1}  of the quantum equilibrium distribution $\mu^\Psi_B$ on shape space. 
}

\medskip 
In the 3-gauge this can be written in a more familiar and explicit manner:
\begin{align}
\Pconf^{\Psi_0}(X_t\in dx\, | Y_t= Y) &= 
 | \liftw{\psi}_{3, t}(\lift{x})|^2 d \lift{x} \\ &= C
 |\widehat{\Psi}_{3 , t}(\lift{x}, \lift{Y} )|^2 d \lift{x}
\,,
\end{align}
 where  $ \liftw{\psi}_{3, t}$ is the (normalized) conditional wave function in the 3-gauge in the frame $\lift{Y}$ and $C$ is a normalization constant.

\subsection{The Physical Significance of the Conditional Distribution for Stationary Wave Functions}
\label{FCPSW}
However, what has just been said is not quite right for  the \Sc gauge corresponding to the transition from \eqref{eq:Steptwo} to \eqref{step312}. As already indicated at the end of Sect.  \ref{rmkbm}, this involves no change of measure for self-adjointness, so that   $\mu^{\zerogaupsi} \neq \mu^{\gaupsi}$. This might seem bad. On the other hand, the change in $ \mu^{\gaupsi}$ is precisely the one implied by the random time change arising from replacing $f^{-1}\nablab$ by $\nablab$, as described above \eqref{bohmabsolu}. This seems sort of good. But one should be puzzled by the fact that this random time change leads to a change in the measure $ \mu^{\gaupsi}$, which would seem to have some physical significance.
But how could it, since the random time change has no physical significance? 

So there are several  questions here that need to be understood better: (i) What is genuinely physically significant in  $ \mu^{\gaupsi}$? (ii) How does that resolve the apparent problem that in the \Sc gauge we are dealing with a $\mu^{\zerogaupsi}$ that is incompatible with $\mu^\Psi_B$ and thus apparently one that would yield an incompatible fundamental conditional probability formula in the \Sc gauge?

A crucial ingredient in an  answer to these questions is the claim that it is not exactly  $ \mu^{\gaupsi}$, resp.  $ \mu^{\gaupsi_S}$,  that is physically relevant. Rather what is physically relevant   is the associated current $J^{\gaupsi}=\rho^{\gaupsi} v^{\gaupsi}$,   resp. $J^{\gaupsi_S}=\rho^{\gaupsi_S} v^{\gaupsi_S}$,  where $\rho^{\gaupsi} \propto \rho^{\gaupsi_3}$,
resp. $\rho^{\gaupsi_S} $,  is  the density
of  $\mu^ {\gaupsi}$,  resp.  $\mu^ {\gaupsi_S}$, with respect to the Lebesgue measure.  The current  is invariant under all the transitions, either because both factors are or because the changes in the factors compensate each other.

Now why should the current be what is physically relevant? Because it yields the same crossing probabilities for hypersurfaces---that either yield the probability distribution on geometrical paths in configuration space, or the probability distribution associated with the return map for Poincar\'e sections corresponding to physical situations on which we wish to condition. Concerning the former, this could correspond to conditioning on the value of a suitable clock variable, for which the corresponding conditional wave functions have more familiar quantum evolutions, so that first conditioning on such a clock variable would put us back in a more familiar situation to which the argument described above would apply.

{
Be that as it may, let's return to the question of why the  change in the measure  $ \mu^{\gaupsi}$  ($\mu^{\zerogaupsi} = f^{-1}\mu^{\gaupsi}$) arising from the random time change has no physical significance. We have argued that the main physical relevance of  $ \mu^{\gaupsi}$  resides in the implied conditional distribution of a subsystem given its environment. Suppose $f$ depends only upon the environment. In this case, the change in measure associated with $f$ produces no change in the corresponding conditional distribution.\footnote{Letting $\lift{x}$ and $\lift{y}$ be the configuration variables of system and environment, respectively, note that if $f$ does not depend on $\lift{x}$, the  kinetic energy term   in the Hamiltonian  \eqref{exvtwoxx} will be 
$$
-\frac{\hbar^2}{2} \left[  \frac{1}{f} ( \nablab^2 )_{\lift{x}}+  
 ( \nablab\bigcdot \frac{1}{f}\nablab )_{\lift{y}}  \right] \,,
$$
where 
$\nablab $ and  $\nablab\bigcdot$ are the mass-weighted Euclidean gradient and divergence and the subscripts refer  to their restrictions to the $\lift{x}$-variables and the $\lift{y}$-variables, respectively. Accordingly, whenever the system is decoupled from its environment, its conditional wave function $\liftw{\psi}_{3} (\lift{x}) =  \liftw{\Psi}_{3}(\lift{x}, \lift{Y}) $ will evolve (after suitable  rescaling) according to the standard Schr\"odinger Hamiltonian with  masses $ f (\lift{Y})  m_\alpha$, $\alpha = 1, \ldots M$, where $\lift{Y}$  is the actual configuration of the environment. Thus,
in this case,    the effect of the environment on the system corresponds just to a (possibly time-dependent) rescaling of the masses. Then it turns out that the  3-gauge is  more similar to the Schr\"odinger gauge than one may have expected.}
  And it seems likely that for reasonable choices of $f$, such as those given {in Sect.\ref{conformalfactors}} above, it will approximately be a function of the environment, with  negligible error for subsystems of reasonable size, much smaller than that of the universe.

But even if this is so, the question remains as to exactly what of physical significance this conditional distribution represents. 
After all,   the transition to the \Sc gauge required relational time, but if we take relational time seriously,
%
%
what is physical is not the configuration $Q_t$  of the universe at some time $t$, but the geometrical path of the full history of the configuration, with no special association of the configurations along a path with times.
In this (more physical) framework, the conditional distribution of the configuration $X_t$ of a subsystem given the configuration $Y_t$ of its environment is not meaningful.


What is meaningful is (i) a probability distribution  $\Ppath$ on the space $\mathscr{P}$ of (geometrical, i.e. unparametrized) paths (determined by the current, for example by using a cross-section, as we shall explain in  Sect. \ref{pathmeasures}) 
and (ii) the conditional distribution relative to $\Ppath$ of the configuration $X_Y$ of the subsystem when the path $\gamma\in \mathscr{P} $ has environmental configuration $Y,$ {\it given} that the path passes through a configuration with environment $Y$, $Y \in \gamma$.\footnote{This seems to involve a new sort of conditional probability analogous to the so-called Palm measure. In particular, it does not appear to be a special case of the conditional distribution relative to  a $\Sigma$-algebra or a foliation or given the value of another random variable. That is because any path will involve configurations with many different environments $Y.$} We assume  here 
that there is at most one such configuration for (any) $Y$.\footnote{When $\Psi_t$ is time-dependent, it is natural to suppose that the time-parameter $t$ has physical significance via the changes in typical configurations arising from changes in $\mu^{\Psi_t}$. In this situation, treating time as if it were physical and observable seems to be a reasonable approximation. In any case, this is an approximation we almost always make, and it seems to often work quite well.}

\subsection{The Association Between Measures on Path Space and  on Configuration Space}
\label{pathmeasures}
We shall now elaborate on the relations between
measures on    a configuration space $\mathscr{Q}$  and measures on  a set  $\mathscr{P}$ of (smooth) paths  on  $\mathscr{Q}$ forming a one-dimensional foliation of  $\mathscr{Q}$.  
%
%
%
%
%
First, let us fix the notations: 
we shall denote by $\gamma$ a path in  $\mathscr{P}$, by  $q$   a  point in  $\mathscr{Q}$  and by  $\gamma (q)$ the  path passing  through~$q$. 

We introduce  the notion of 
{\em time function} on  $\mathscr{Q}$: a   smooth real-valued function   $\tau=\tau (q)$,  $q\in \mathscr{Q}$,   which is monotonic   (and extends from $-\infty$ to $+\infty$) on each path $\gamma \in \mathscr{P}$. The latter is related to, but different from,  the  notion of {\em dynamics} on $\mathscr{Q}$ (in the sense of  the theory of dynamical systems): 
a one parameter family of (invertible and smooth) maps  $T=T_s$ on  $\mathscr{Q}$,  with time $s$ varying on the reals.  Clearly,  a 
time function $\tau$ generates a dynamics  $T=T^\tau$  such that
\begin{equation}\label{compat}
\tau (T_s (q)) = \tau (q) + s\,.
\end{equation}

Let  $\tau$ be  a time function and 
%
denote by
$d\tau_\gamma (q)$   the infinitesimal increment of time (with respect to the time function $\tau$)
along a path $\gamma$ passing through $q$. 
Consider the function on path space (random variable)
giving the time spent in the region $A\subset \mathscr{Q}$ by the path, relative to the time function $\tau$:
\[
\tau (A) = \tau_\gamma (A) = \int_A d \tau_\gamma (q)\,.
\]
Let \begin{align}\label{exptaua}
\Epath \left(\tau(A)\right) =\int_{\mathscr{P}}  
 \tau_\gamma (A)   \Ppath(d\gamma)  
\end{align}
be  the expected value with respect to $\Ppath$ of $\tau (A)$ (i.e., the expected time spent by the path in $A$)  {\em relative to the time function $\tau$}.
We say that  {\em   $\mu$ is associated  with  $\Ppath$    and  $\tau$ if }
\begin{align}
\mu(A)=\Epath \left(\tau(A)\right) \,.
\end{align}

The measure $\mu$ can also be described as follows. We may identify configuration space $\mathscr{Q}$ with $\mathscr{P}\times \mathbb{R}$ by associating any configuration $q$ with the path to which it belongs and the time along the path:
$$  q \mapsto (\gamma(q), \tau (q) ) \,. $$
Under this identification $\mu$ is the product of $\Ppath$ and the Lebesgue measure, $d\mu = d \Ppath\, dt $.

If  $\mu$  is associated with $\Ppath$  and $\tau$, then  we have that $ \Ppath$ arises from  conditioning  $\mu$ on  $\tau$, 
\begin{align} 
\label{probpathcon}   \Ppath (S) =     \mu(\check{S}\, | \, \tau=t)  \end{align}
where $S$ is a set of path and $\check{S}=\{ q\in \mathscr{Q}\,  | \gamma(q) \in S \} $. Note that the right hand side of \eqref{probpathcon}  does not depend on $t$.
(It follows from \eqref{compat} and  \eqref{exptaua} that  $\mu$ is stationary with respect to the dynamics $T_s$ generated by $\tau$ according to \eqref{compat}.) Conversely,  when $\mu$, $\Ppath$, and $\tau$ are related as in \eqref{probpathcon},   $\mu$ is associated with $\Ppath$ and $\tau$.

Suppose now  that $\mu$ is stationary for {\em a dynamics $T$ for which $\tau$ is a time function},\footnote{Note that, as a consequence   of Poincar\'e  recurrence, 
  in general  there might be no  time function associated with  a dynamics $T$.  However,  if one does exist, it cannot be  unique, because it would depend on the choice of an initial cross section.}
 i.e., such that \eqref{compat} is satisfied.  Let $\Ppath$ be a measure on $\mathscr{P}$, the path   space of the dynamics. Then the following are equivalent:
\begin{subequations}
\begin{align}
\Ppath (S) &= \mu \left( \left\{ q\in\mathscr{Q}\, | 0\le \tau(q) \le 1 \;\text{ and }\;
\gamma(q)\in S   \right\}  \right)\\
\Ppath (S) &=   \mu(\check{S}\,| \, \tau = t) \,.
\end{align}
\end{subequations}
When $\mu$ and $\Ppath$  are associated in this way  then say that {\em  $\Ppath$ is generated by $\mu$ and $T$}, and if this is so, then $\mu$ is associated with $\Ppath$  and $\tau$.
In particular, if $\tau$ is a time function for  any Bohmian dynamics with stationary $\Psi$ (or $\mu^\Psi$) then $\mu^\Psi$ is associated with $\Ppath^\Psi$ and $\tau$, where $\Ppath^\Psi$  is the measure generated by  $\mu^\Psi$  and the Bohmian dynamics. 

Furthermore,  if  the  dynamics $T$ is  given by a vector field $v$, $\Ppath$ is generated by $\mu$  and   $T$ if and only if 
\begin{align}
\label{probpathcur} \Ppath (S)  =  \int _{\Gamma_S} J   \bigcdot d\sigma\,,
\end{align}
where  $J= J(\mu, T)$ is the current associated with the dynamics. Here  $\Gamma_S = \Gamma  \cap \check{S}$, where  $\Gamma$ is any  cross section of the foliation  $ \mathscr{P}$,  for example  the level surface  $\tau (q) =  t  $. If  $\mu$ is absolutely continuous with density $\rho$ (with respect to the coordinate  measure in  some  coordinate system), then  $J= v \rho$  (with $v$    represented in that  coordinate system) and the surface integral above is given by the usual formula. 

%
%
%

\subsection{ A Conditional Probability Formula for Path Space Measures}

In  this section  and the next we consider a fixed system (and its environment), with configuration $Q=(X,Y)$. We also fix the configuration $Y$ of the environment.
We   prove    a  crucial fact for establishing  in Sect. \ref{fucofor} the fundamental conditional probability formula for stationary wave functions, a general    fact that  we call the  {\em  path space conditional probability formula}.

In this  formula we condition on the path space event $\{ Y\in \gamma\} \subset \mathscr{P}$ that the path $\gamma$ contains 
a configuration with environmental configuration $Y$, i.e., on the set of paths $\gamma$ that pass through a configuration with environment $Y$.  And in this formula we are interested in the conditional distribution of $X_Y$, the configuration of the system when the path 
 passes through the  configuration with environment $Y$, a function (random variable)  on the subset 
 $\{ Y\in \gamma\}$ of path  space $\mathscr{P}$.

\medskip 
\noindent{\bf Path space conditional probability formula:}
{ {\em  Let $\Ppath$ be a  measure  on  the path space $\mathscr{P}$. Suppose $\mu$ is  the measure on  the corresponding configuration space 
$ \mathscr{Q}$  associated with  $\Ppath$ and a time function $\tau$ that depends only upon the configuration  $Y$ of the
 environment.    Then
 \begin{equation}\label{paspconfor}
\Ppath(X_Y\in dx|Y\in\gamma) = \mu(X\in dx|Y) \,.
\end{equation} }}
\medskip

This follows more or less  as before (see the second bullet of Sect. \ref{fundaconform})  with  $\mathscr{E}$,  as before,  corresponding to the  configuration of the environment and now $\mathscr{F}$  corresponding to the foliation given by the time function. In more detail, 
since the time function depends only on $Y$, 
$\mathscr{E} \subset \mathscr{F}$.   
Under the natural  identification of path space with any leaf of  $\mathscr{F}$,  the event $\{X_Y\in dx\}$ becomes $\{ X\in dx\}$, the event  $\{ Y\in \gamma\}$ becomes the  event that the environment has configuration $Y$, and 
by \eqref{probpathcon}   $\Ppath$   becomes  the appropriate conditional measure.  

\subsection{The Fundamental Conditional Probability Formula for Stationary  Wave Functions}\label{fucofor}

We   assume now  that there exists a time function for the Bohmian dynamics that depends only on the configuration of the environment, an assumption that we  call {\em the existence of   clock variables 
for the Bohmian  dynamics}.
Under this assumption
%
%
the conditional distribution  $\Ppath^\Psi(X_Y\in dx|Y\in\gamma) $ 
is given by the usual Born's rule on absolute configuration space.

\medskip 
\noindent{\bf Fundamental conditional probability formula for stationary wave functions:}  {\it Suppose that there   exists a clock variable for the Bohmian dynamics generated by a stationary wave function $\Psi$ on shape space, as well as one for the Bohmian dynamics in the Schr\"odinger gauge. Then  }
\begin{equation}\label{cp2}
\Ppath^\Psi(X_Y\in dx|Y\in\gamma)  = \mu^{\widehat{\Psi}} (\lift{X}\in d\lift{x}|\lift{Y} )
\end{equation} 
{\it in any of the first three gauges,}\footnote{These formulas, for the different gauges, may appear to be incompatible. But for the condition relating $\Psi$ and  the time-parameter to hold for both the \Sc gauge and any   other gauge, the condition on $f$ mentioned above would presumably have to be satisfied, in which case the formulas would agree.}  or \begin{equation}\label{cp1}
\Ppath^\Psi(X_Y\in dx|Y\in\gamma)  =    C |\zerogaupsi(\lift{x},\lift{Y})|^2d\lift{x} 
\end{equation}
{\it in the \Sc gauge (as before, $C$ is a normalization constant) }
{\it where $\lift{Y}$  is   a lift of $Y$ and $\lift{x}$ is the lift of $x$ in the frame $\lift{Y}$. Here $\widehat{\Psi}$ and  $\widehat{\Psi}_S$ are the lifts of $\Psi$ in any of the first three gauges  or  in the Schr\"odinger gauge, respectively. }
\medskip

This is  
a consequence of the   path space probability formula \eqref{paspconfor}.
Applying  it  to the Bohmian dynamics  (i) on shape space, (ii) in the first three gauges, and (iii) in the Schr\"odinger gauge, we obtain the following:
\begin{align}\label{bas1}
\Ppath^\Psi(X_Y\in dx|Y\in \gamma) &= \mu_B^\Psi (X\in dx|Y)  \\
\label{bas2}
\Ppath^{\widehat{\Psi}}(\lift{X}_{\lift{Y}} \in d\lift{x}|\lift{Y} \in \boldsymbol{\gamma}) &= \mu^{\widehat{\Psi}} (\lift{X}\in d\lift{x}|\lift{Y}) \\
\label{bas3}
\Ppath^ {\zerogaupsi} (\lift{X}_{\lift{Y}} \in d \lift{x}|\lift{Y}\in  \lift{\gamma}) &= \mu^ {\zerogaupsi} (\lift{X}\in d\lift{x}|\lift{Y})\,,
\end{align} 
given the existence of clock variables for the Bohmian dynamics. Here   $\boldsymbol{\gamma}$ is  a path   for the dynamics on  absolute configuration space.

By \eqref{last1}  the right hand sides  of  \eqref{bas1} and   \eqref{bas2}  agree when  $\lift{x}$ and $\lift{Y}$ are   appropriate  lifts of $x$ and $Y$, so that \eqref{cp2} follows.  Since
the currents $J$ associated with  $ \mu^{\gaupsi}$ and  $ \mu^{\gaupsi_S}$ arising  from their respective  dynamics are the same, it follows that
$\Ppath^{\widehat{\Psi}} = \Ppath^ {\zerogaupsi} $,  so that the right hand sides of 
\eqref{bas2} and  \eqref{bas3}  are also the same. Thus 
\eqref{cp1} follows as well.


%
}

\subsection{Typicality}

As a partial summary, we find that on the absolute configuration space level the dynamics and the probabilities for subsystems should be of the usual form. While it is true that on the universal level the connection between $|\Psi|^2$ and probability, or, more precisely, typicality, would be broken, this would not be visible in any of the familiar every day applications of quantum mechanics, which are concerned only with subsystems and not with the entire universe.

In particular the patterns described by the quantum equilibrium hypothesis will be typical with respect to a measure,  not on absolute configuration space, but on shape space, on the fundamental level, which is fine. There is a widespread misconception with respect to Bohmian mechanics that $|\Psi|^2$ for the universe and $|\psi|^2$ for subsystems play, physically and conceptually, similar roles. They do not, since the role of $|\Psi|^2$ is typicality while that of  $|\psi|^2$ is probability. If this distinction is too subtle, the fact that, from a relational perspective, these objects live on entirely different levels of description, $|\Psi|^2$ on the fundamental level, i.e., on shape space,  and $|\psi|^2$ on absolute configuration space, might make it easier to appreciate how very different they are.
}

\section{Outlook} 
{The basic problem in cosmology is to determine  which laws govern the universe as a whole. The traditional approach is that of building  a story about the universe  starting from  the physical laws  operating at small scales, such as the Standard Model of particle physics, and  incorporating them within a theory containing a  now missing quantum theory of gravity. However, the relational point of view suggests that there is something basically wrong in treating the universe as a whole as a mere combination of  the  systems that compose it, say galaxies or cluster of galaxies. We elaborate.}

{Shape space physics is genuinely holistic, and suggests the holistic character of quantum physics associated with entanglement and quantum nonlocality.
To appreciate this point, note that for relational space the state of the universe at a particular location is not, in and of itself, meaningful. In that sense, for shape space physics, there are no local beables, so that locality itself can't  easily be meaningfully formulated. Similarly one can't meaningfully consider the behavior of individual particles without reference to other particles, since there is no absolute space in which an individual particle could be regarded as moving. And even for a pair of particles, to speak meaningfully of the distance between them, a third particle would be required, to establish a scale of distance. And similarly for galaxies.}

{There is one rather conspicuous relational aspect that we've ignored. For indistinguishable particles we should have taken one further quotient and enlarged the similarity group $G$ to include the relevant permutations of particle labels. We believe that this would not be too difficult to do, but have chosen not to do so here.}

{Quite a bit more difficult is the connection between relational physics and relativistic physics.}

\begin{itemize}

 \item {A simple point: In relational physics as discussed here the traditional separation of space and time is retained. While configuration space is replaced by shape space, and time becomes non-metrical,  shape space  retains an identity separate and distinct from that of (non-metrical)  time. This is in obvious contrast with relativistic physics, in which space and time lose their separate identities and are merged into a space-time.}

 \item {{\it Simultaneity regained and simultaneity lost:} Perhaps the most characteristic feature of relativity is the absence of absolute simultaneity. Not so for relational physics. Since it retains the separation of space and time,  an absolute simultaneity is built into the very structure of relational physics as described here. Nonetheless, there is a sense in which simultaneity is lost. As discussed in  Sect.\ref{FCPSW}, with relational time the notion of the configuration (or shape) $Q_t$ of the universe at ``time $t$'' is not physically meaningful. And with what is meaningful---geometrical paths in the space of possible configurations (or shapes)---one can no longer meaningfully compare or ask about the configurations for two different possible histories at the same time. Given the actual configuration of the universe, it is not meaningful to ask about the configuration of an alternative history at that time without further specification of exactly what that should mean.}

 \item {Can the relational point of view be merged with or extended to relativity? Can we achieve a  relational understanding of space-time? General relativity is certainly a step in that direction, but it does not get us there. Space-time in general relativity is metrical---in a way that neither space nor time are in relational physics. A complete extension, if at all possible, is a real challenge.}

 \item {Another possibility: relativity is not fundamental, but---like Newtonian physics in the Newton gauge  and quantum physics in the Schr\"odinger gauge---is, instead,  a consequence of a suitable choice of gauge. This possibility, which is suggested by the work of Bryce DeWitt \cite{dewitt1970spacetime} and Barbour and {coworkers} {(see, e.g.,   \cite{barbour2002relativity},  \cite{gomes2011einstein}, and \cite{barbour2012shape})}, would be worth carefully exploring.}

\end{itemize}

The discussion in  Sect. \ref{FCPSW}, with its focus on geometrical paths as more fundamental from the point of view of relational time (and, more generally, with regard to  what is more directly observable),  was based on a Bohmian approach to quantum physics. This approach involves a law for the evolution of configurations, yielding geometrical paths, the analysis of which leads ultimately to the Born rule (on absolute space) in a more or less familiar form. Without such an approach---and the paths that it provides---it is not easy to see how one could begin to proceed in a principled manner.

As is well known,  non-normalizable wave functions  tend to occur in quantum cosmology. Such wave functions would normally be regarded as problematical and unphysical (since the formal structures of orthodox quantum mechanics, with their associated probabilities, are crucially based on the notion of a Hilbert space of square-integrable, i.e. normalizable, wave functions).

However, for our analysis starting in Sect.  \ref{FCPSW}  the connection between a measure on path space and a  (stationary) non-normalizable measure on configuration space turned out to be crucial. By the very nature of this connection   the measure on configuration space and its associated wave function in fact  had to be  non-normalizable.
Hence what  from an orthodox perspective is a vice is transformed into a virtue in relational Bohmian mechanics.
\begin{acknowledgments}
We are grateful  to  Florian Hoffmann for his  input to a very early draft of this paper and to  Antonio Vassallo for his insights. We thank Sahand Tokasi for stimulating discussions.  We thank Eddy Chen and Roderich Tumulka  for a careful reading of the manuscript and useful suggestions. The many discussions with Julian Barbour are gratefully acknowledged, especially for sharing with us in his well known  enthusiastic way his ideas on shape dynamics. N. Zangh\'{\i} was supported in part by INFN.
\end{acknowledgments}


\section*{Appendix: Some facts about second-order partial differential operators}\label{App}
\addtocontents{Appendix: Some facts about second-order partial differential operators}

In local coordinates,  any   second order  partial differential operator (PDO)
with real coefficients, self-adjoint with respect to some volume element $\mu(dq)$  is of the form
\begin{align}\label{transrhpop}
\Op=\sum_{ij} A^{ij}\partial_i \partial_j + \sum_i B^i\partial_i + \C \end{align}  with symmetric matrix function $A= (A^{ij} ) $,  vector  $B =(B^i) $, and scalar $\C$. 
So it can be compactly written as
\begin{align}\label{transrhpop1}
\Op = A\nabla \nabla + B \cdot \nabla +\C\,.
\end{align}
Note that, while the explicit functions $A= A(q)$, $B= B(q)$ and $\C=\C(q)$ depend on the coordinate system chosen,  $A$ is in fact a tensor, so that the fact that two second order PDOs $L$ and $L'$ have equal $A$-parts  (pure second-derivative parts) is invariant.

Note that for a Laplace-Beltrami operator $\C=0$.  Moreover,  we shall need the $A$-part  
of the Laplace-Beltrami operator  with respect to the invariant metric $g= f g_e$, where $g_e$ is the mass-weighted  Euclidean metric.  According to \eqref{eq:LB}, we have  \[ A= { f ^{-1}} I \,,\]
where $I$ is the identity matrix. 

Here are some relevant facts:
Suppose $\Op$ and $\Op ' $ are  second order  PDOs on a manifold $M$. If 
\begin{enumerate} 
\item  they have the same $A$-part, and
\item are self-adjoint with respect the same measure $\mu$,
\end{enumerate}
then they differ by at most a multiplication operator $D=D(q)$, i.e.,
\[ \Op' = \Op+  D\,.\]  This is so  because their difference, which must be of the form $B\cdot \nabla + D$, must also be self adjoint. Since $D$ is as well, $B\cdot \nabla$  must also be. But for no measure $\mu$ can $B\cdot \nabla$  be self-adjoint on $L^2(d\mu)$, unless $B=0$. Moreover, if $\Op$ has no $\C$-part, then \[ D = \Op' 1 \,,\]
where $1$ is the constant function  equal to $1$.

%

\end{document}